\PassOptionsToPackage{dvipsnames}{xcolor}

\PassOptionsToPackage{hyperfootnotes=false}{hyperref}
\documentclass[published]{agujournal2025}
\usepackage{url} %
\usepackage{lineno}
\usepackage{soul}
\usepackage{booktabs}       %
\usepackage{placeins}
\usepackage{etoolbox}
\usepackage{longtable}      %
\usepackage{nicefrac}       %
\usepackage{microtype}      %
\usepackage{graphicx}
\usepackage{pstool}
\usepackage{amsmath}

\usepackage{amsbsy}
\usepackage{bm}
\usepackage[mathscr]{eucal}
\usepackage[font={footnotesize}]{caption}
\newcommand{\footnotestyle}{}

\newcommand{\code}[1]{\texttt{#1}}

\usepackage[charsperline = 90,
            autoload = true,
            defaultmonofont = false,
            linenumbers = true,
            theme = default]{jlcode}
\renewcommand{\jlbasicfont}{\ttfamily\fontsize{7.0588235}{8}\selectfont}
\calcboxlengths

\definecolor{lightgray}{gray}{0.97}
\definecolor{darkgray}{gray}{0.45}

\usepackage{scalerel, stackengine}
\journalname{Journal of Advances in Modeling Earth Systems}
\hypersetup{breaklinks = true,
	        colorlinks = true,
	        linkcolor  = blue,
            urlcolor = RoyalPurple,
	        citecolor  = Blue}

\stackMath
\newcommand\tsup[2][2]{%
 \def\useanchorwidth{T}%
  \ifnum#1>1%
    \stackon[-.5pt]{\tsup[\numexpr#1-1\relax]{#2}}{\scriptscriptstyle\sim}%
  \else%
    \stackon[.5pt]{#2}{\scriptscriptstyle\sim}%
  \fi%
}

\setstackEOL{\#}
\stackMath
\def\hatgap{2pt}
\def\subdown{-2pt}
\newcommand\reallywidehat[2][]{ \renewcommand\stackalignment{l} \stackon[\hatgap]{#2}{ \stretchto{
    \scalerel*[\widthof{$#2$}]{\kern-.6pt\bigwedge\kern-.6pt}
    {\rule[-\textheight/2]{1ex}{\textheight}}}
    {0.5ex}_{\smash{ \belowbaseline[\subdown]{\scriptstyle#1} }}
}}

\newcommand*\mystrut[1]{\vrule width0pt height0pt depth#1\relax}
\renewcommand{\strut}{\mystrut{1ex}}
\newcommand{\inlinejll}{\vspace{-2ex} \begin{jllisting}}

\renewcommand{\arraystretch}{1.5}
\renewcommand{\b}[1]    {\symbfit{#1}}
\renewcommand{\r}[1]    {\mathrm{#1}}
\renewcommand{\c}       {\, ,}
\renewcommand{\d}       {\partial}

\newcommand{\bu}        {\symbfit u}
\newcommand{\bU}        {\symbfit{U}}

\newcommand{\buS}       {\symbfit u_s}
\newcommand{\bubg}      {\bu_b}
\newcommand{\bubgc}     {\bu_{p}}
\newcommand{\ubg}       {u_b}
\newcommand{\cbg}       {c_b}

\newcommand{\bx}        {\b{x}}
\newcommand{\bxh}       {\hspace{0.1em} \symbfit{\hat x}}
\newcommand{\byh}       {\hspace{0.1em} \symbfit{\hat y}}
\newcommand{\bzh}       {\hspace{0.1em} \symbfit{\hat z}}
\newcommand{\bnh}       {\hspace{0.1em} \symbfit{\hat n}}

\newcommand{\ee}        {\r{e}}

\newcommand{\dd}        {\r{d}} 
\newcommand{\di}        {{\, \dd}}

\newcommand{\half}      {\tfrac{1}{2}}
\newcommand{\beq}       {\begin{equation}}
\newcommand{\eeq}       {\end{equation}}
\newcommand{\beqs}      {\begin{gather}}
\newcommand{\eeqs}      {\end{gather}}

\newcommand{\defn}      {\ensuremath{\stackrel{\r{def}}{=}}}

\newcommand{\bnabla}    {\b{\nabla}}
\newcommand{\nablah}    {\bnabla_{\! h}} 
\renewcommand{\div}     {\bnabla \bcdot}

\newcommand{\bcdot}     {\b{\cdot}}

\newcommand{\com}       {\, ,}
\newcommand{\per}       {\, .}

\begin{document}
\raggedright

\title{High-level, efficient ocean modeling at all scales with Oceananigans}

\authors{Gregory~L.~Wagner\affil{1, 2},
         Simone~Silvestri\affil{1, 3},
         Navid~C.~Constantinou\affil{4,5},
         Ali~Ramadhan\affil{6},
         Jean-Michel~Campin\affil{1},
         Chris~Hill\affil{1},
         Tom\'as~Chor\affil{6,7},
         Jago~Strong-Wright\affil{8},
         Xin~Kai~Lee\affil{1},
         Francis~Poulin\affil{9},   
         Andre~Souza\affil{1},
         Keaton~J.~Burns\affil{1,10},
         Siddhartha~Bishnu\affil{1,8},
         John~Marshall\affil{1}, and
         Raffaele~Ferrari\affil{1}}

\affiliation{1}{Massachusetts Institute of Technology, Cambridge, MA, USA}
\affiliation{2}{Aeolus Labs, San Francisco, CA, USA}
\affiliation{3}{Politecnico di Torino, Torino, Italy}
\affiliation{4}{University of Melbourne, Parkville, VIC, Australia}
\affiliation{5}{Australian Research Council Center of Excellence for the Weather of the 21st Century, Australia}
\affiliation{6}{atdepth MRV, Cambridge, MA, USA}
\affiliation{7}{University of Maryland, College Park, MD, USA}
\affiliation{8}{University of Cambridge, Cambridge, United Kingdom}
\affiliation{9}{University of Waterloo, Waterloo, ON, Canada}
\affiliation{10}{Flatiron Institute, New York, NY, USA}
\date{\today}

\keypoints
{Oceananigans combines a reusable finite-volume framework and programmable interface for simulating oceanic motion at all scales.}
{A high-level interface, basic numerics, and GPU acceleration yield accessible code capable of high-fidelity simulations.}
{Combining accessibility with state-of-the-art performance accelerates model development, and thus the progress of Earth system science.}
\authoraddr{Gregory L. Wagner}
\citation{}
\maketitle

\begin{abstract}

We describe the user interface, governing equations, and numerical methods underpinning the ocean modeling software called ``Oceananigans''.
Oceananigans development has been led by the Climate Modeling Alliance as part of a larger effort to build a trainable climate model with quantifiable uncertainty.
Notably, Oceananigans is written in the Julia programming language, is optimized for GPUs, and can efficiently simulate all scales of ocean motion ranging from millimeter-scale turbulence in a small box to planetary-scale ocean circulation.
We argue that a reusable finite-volume framework on structured staggered grids, GPU-efficient algorithms, and accessible interfaces and source code together enable efficient all-scale simulations and rapid innovation in model development.
This design negotiates a dual mandate for highest-possible performance (to support state-of-the-art applications) and enhanced accessibility (to facilitate adoption and development).
The dual mandate aims ultimately to accelerate the progress of Earth system science.
\begin{plainlanguagesummary}
This paper describes Oceananigans, a tool for simulating ocean currents and fluid motion written in the Julia programming language.
Using a relatively new programming language separates Oceananigans from usual software written in Fortran or C++ and places it alongside a handful of recent efforts to modernize Earth system modeling software.
Oceananigans can simulate ocean motion on a wide range of scales from millimeters to planetary-scale.
Oceananigans is very fast because it is optimized for graphics processing units (GPUs), which comprise most of the biggest computers today.
We argue that the Oceananigans design strategy, which combines basic numerics on GPUs with a powerful user interface, can accelerate the pace of model development and therefore progress in Earth system science.
\end{plainlanguagesummary}
\end{abstract}

\section{Introduction}

Only numerical models can entirely synthesize the vast accumulation of knowledge about the ocean and Earth system.
As a result, the capabilities of modeling software rate-limit progress in Earth system science.
Since the first general circulation models ran on primitive computers \cite{phillips1956general, bryan1969numerical}, advances in hardware, numerical methods, and the approximate parameterization of otherwise unresolved processes have improved the fidelity of ocean simulations \cite{griffies2015historical}.
But the gap between potential and practice in ocean modeling is widening because advances in software and hardware technology have outpaced innovation in model development.
Today, most ocean modeling software
\textit{(i)} can tackle only a limited subset of ocean modeling problems,
\textit{(ii)} does not run on the fastest computers (most of which are based on GPUs),
\textit{(iii)} has software interfaces and architectures that make it difficult to share and reuse developments across applications.

This paper describes the design of Oceananigans, ocean modeling software written in the Julia programming language \cite{bezanson2017julia}.
Oceananigans development was initiated by the Climate Modeling Alliance as part of a larger effort to develop a climate model automatically-calibrated to observations and targeted large eddy simulations, and with quantified uncertainty \cite{schneider2017earth}.
Our ambition to support both research and forecasting is inspired by the success of the Weather Research and Forecasting model (WRF) \cite{skamarock2019wrf}.
Oceananigans proposes to close the gap between potential and practice in ocean modeling with three interlocking design elements:
\textit{(i)}~a coherent finite-volume framework on structured orthogonal grids, with shared numerical operators supporting the full range of ocean-flavored fluid dynamics from millimeter-scale direct numerical simulation, through meter-scale large eddy simulation, to planetary-scale ocean circulation,
\textit{(ii)}~an algorithm optimized for high performance on GPUs that yields significant performance gains over existing CPU-based codes, and
\textit{(iii)}~a programmable, script-based user interface that leverages modern language features for configuring, running, and analyzing simulations.
In this way, Oceananigans is a modeling system for \textit{all scales}, designed to enable low-cost \textit{efficient} high resolution simulations, and controlled with a \textit{high level} interface optimized for productivity.
The goal of this paper is not to report on all aspects of the model, which is rapidly evolving, but rather to describe, evidence, and explain how each of these design elements is allowing fast progress in both the model development and the science it allows. We demonstrate their cooperative utility through a series of examples that progress from idealized two-dimensional turbulence to realistic global ocean simulations.

\subsection{Interlocking design elements: fast, flexible, and friendly}

The evolution of global ocean circulation over millennia is controlled by turbulent mixing with scales that range down to seconds and millimeters.
Two distinct systems have evolved to model this huge range of oceanic motion: ``GCMs'' (general circulation models) for hydrostatic regional-to-global scale simulations, and simpler software for nonhydrostatic large eddy simulations (LESs) with meter-scale grid spacing that are high-fidelity but limited in duration and extent.
Compared to LESs, GCMs usually invoke more elaborate numerical methods and parameterizations to cope with the global ocean's complex geometry and the more significant impacts of unresolved subgrid processes.
Oceananigans bridges this divide: beginning as software for LES \cite{ramadhan2020oceananigans} by refining an approach for hybrid hydrostatic/nonhydrostatic dynamical cores pioneered by MITgcm \cite{marshall1997finite}, Oceananigans has grown into a single framework for simulating ocean-flavored fluid dynamics at any scale.

A structured finite volume algorithm on C-grids, optimized from the start for GPUs \cite{silvestri2025gpu}, yields significant acceleration over existing CPU-based codes \cite{silvestri2023oceananigans}.
LES-inspired, minimally dissipative advection schemes \cite{silvestri2024new} achieve up to 2x higher effective resolution than existing methods, automatically adapting to local flow conditions and grid spacing.
The resulting modeling system combines regular, structured-grid data layouts, GPU-efficient algorithms, and minimally dissipative advection schemes.
Our working hypothesis is that these features enable a resolution-focused approach to accuracy, in which finer grids \cite{chassignet2017impact, kiss2020access, chassignet2021importance} can offset the need for more complex approaches such as generalized vertical coordinates~\cite{shchepetkin2005regional, leclair2011z, petersen2015evaluation}, Lagrangian vertical advection~\cite{halliwell2004evaluation, griffies2020primer}, or unstructured horizontal grids~\cite{ringler2013multi, danilov2017finite, korn2022icon}.
This hypothesis is under ongoing evaluation.
This design creates a synergy between GPU performance, a resolution-based approach to accuracy, and code reuse across scales: shared data structures and numerical operators support both nonhydrostatic DNS and LES and hydrostatic regional and global ocean simulations.
Detailed performance analysis is provided by~\citeA{silvestri2023oceananigans} and~\citeA{silvestri2025gpu}; automated benchmarks are run as part of the test suite and are published at \href{https://numericalearth.github.io/OceananigansBenchmarks}{\nolinkurl{numericalearth.github.io/OceananigansBenchmarks}}, providing up-to-date performance information as the software, hardware, and compiler infrastructure evolve.

Equal to numerical performance for determining the time required to produce a scientific result is the productivity that scientists achieve when using modeling software.
In 1984, Cox published what we believe is the first description of generalizable ocean modeling software~\cite{cox1984primitive, griffies2015historical}, featuring a multi-step workflow of source code modifications followed by namelist configuration.
With more than forty years of progress in software engineering and more than a billion times more computational power, most of today's ocean models bear little resemblance to the Cox model --- \textit{except}, perhaps, for their user interfaces.
Oceananigans implements a programmable, script-first user interface for configuring, running, monitoring, and analyzing simulations, inspired primarily by Dedalus \cite{burns2020dedalus}, with similar approaches taken by SpeedyWeather \cite{klower2024speedyweather} and Thetis \cite{rathgeber2016firedrake, karna2018thetis}.
A central argument of this paper is that such programmable interfaces --- which, unlike text-based namelists, provide a seamless one-step workflow for setup, execution, analysis, and visualization \cite{perez2007iPython, Hunter:2007, paszke2019pytorch, alnaes2015fenics, rathgeber2016firedrake, burns2020dedalus} --- are effective for accelerating progress in Earth system science.
Programmable interfaces facilitate fast prototyping and reproducible simulations, while allowing configurations, diagnostics, and parameterizations to be shared as executable code and reused across applications. They also provide a common interface for scientists and coding agents based on large language models (LLMs) to construct, inspect, and modify simulation workflows.

The use of a modern programming language ties together the different elements of Oceananigans' design.
With Julia, Oceananigans provides a framework wherein new parameterizations \cite{wagner2023catke}, new numerical methods \cite{silvestri2024new}, and new algorithms for performance optimization \cite{silvestri2025gpu} may be both prototyped and refined for production without re-implementation.
Julia's flexibility, expressiveness, and compilation to native machine code enable a wide range of performance optimizations, allowing us to achieve state-of-the-art performance on GPUs.
Structured-grid algorithms reinforce the productivity benefits of a high-level language: regular data structures and composable numerical operators make source code easier to read, reuse, and extend. This extensibility supports companion packages including Breeze for atmosphere modeling, Terrarium for land modeling, ClimaSeaIce for sea ice modeling, and NumericalEarth for coupled simulations.

We do not argue that Julia is strictly superior to other approaches; this kind of framework may also be implemented with domain-specific languages (DSLs) such as Python's JAX \cite{bradbury2018jax} or C++'s AMReX.
Python's main advantage is its status as the \textit{lingua franca} of programming: a much larger open-source community with greater resources devoted to its support and development.
Some of these advantages are mitigated by Julia-Python interoperability and the similarity between Julia and Python syntax.
Both languages enable interactivity, extensibility, automatic installation, and portability to laptops and GPUs \cite{besard2018effective, bradbury2018jax, churavy2024kernel}.
Using a DSL like JAX can have advantages in accelerating development, especially in the initial stages of a project.
At the same time, however, DSLs are restrictive, such that using Julia enables a wider range of strategies for performance optimization comparable to C++ with AMReX without the cost of the ``two-language problem'' and the advantage of flexible model design, algorithm design, and user interface design. Realizing these long-term benefits demands more effort upfront.

The need for next-generation frameworks is evidenced by the recent proliferation of similar efforts.
Regional-mom6 \cite{regional-mom6-JOSS} and CROCO-tools \cite{jullien_2025_15064146} provide Python software that automates the configuration of existing models \cite{adcroft2019gfdl, CROCO-zenodo}, leveraging mature modeling systems with massive user communities and decades of development history, which, in the case of CROCO, support both hydrostatic and nonhydrostatic simulations \cite{auclair2018nonhydrostatic}.
New modeling systems are also being developed in modern programming languages: ClimaAtmos \cite{yatunin2026climate} and ClimaLand \cite{deck2025climaland} provide atmosphere and land surface modeling in Julia; Veros \cite<hydrostatic ocean simulations,>{hafner2021fast, mrozowska2025bayesian} and NeuralGCM \cite<hydrostatic atmosphere simulations,>{kochkov2024neural, yuval2024neural} are written in Python using JAX \cite{bradbury2018jax}; and REMORA \cite{klion2025remora} provides a GPU-capable re-implementation of the Regional Ocean Modeling System (ROMS) built on the AMReX adaptive mesh refinement framework.
JAX-based software benefits from the ubiquity of the Python ecosystem, offers portable performance between CPUs, GPUs, and TPUs, and most notably supports automatic differentiation for gradient-based parameter estimation or training of machine learning components \cite{kochkov2021machine, kochkov2024neural} and adjoint-based data assimilation \cite{solvik20254d}.
Oceananigans achieves performance portability via KernelAbstractions \cite{churavy2024kernel} and differentiability via Enzyme and Reactant \cite{moses2021reverse, moses2026adjoint}, which represent newer alternatives to JAX.

\subsection{Outline of this paper}

The purpose of this paper is to explain Oceananigans' design philosophy and to demonstrate its advantages by providing a sketch of its capabilities, including how its user interface both simplifies basic simulations while enabling complex, creative science.
We note that our goal is not to exhaustively document every closure, feature, or technical detail which are rapidly expanding thanks to a growing number of contributors.
For comprehensive and up-to-date information, we instead refer the reader to the Oceananigans documentation at \href{https://clima.github.io/OceananigansDocumentation}{\nolinkurl{clima.github.io/OceananigansDocumentation}}, which is a living document that evolves with the software.
(Benchmarks may also be found on Oceananigans' \href{https://numericalearth.github.io/OceananigansBenchmarks}{performance dashboard} and also in \citeA{silvestri2023oceananigans},~\citeA{silvestri2025gpu}.)
Forthcoming papers will also describe on-going work, such as the fidelity of global ocean–sea-ice simulations following Ocean Model Intercomparison Project (OMIP) protocols, and regional modeling capabilities and applications.
To that end, this paper proceeds in section~\ref{sec:the-library} by illustrating the basic form of Oceananigans' programmable interface with two classroom examples: two-dimensional turbulence, and a passive tracer advected by two-dimensional turbulence and forced by user-specified forcing.

Section~\ref{sec:equations-parameterizations} continues by writing down the Boussinesq governing equations that underpin Oceananigans' nonhydrostatic and hydrostatic models.
We then sketch out Oceananigans capabilities with a series of examples that progress from basic direct numerical simulations of cabbeling, to realistic tidally-forced nonhydrostatic large eddy simulations over a headland, to a 1/12$^\mathrm{th}$ degree hydrostatic, eddying, global ocean simulation.
To summarize: Oceananigans supports both nonhydrostatic and hydrostatic simulations on rectilinear and curvilinear grids with or without bathymetry.
A suite of pressure solvers, WENO-based advection schemes \cite{shu1997essentially}, Laplacian diffusivity, and subgrid closures including constant and dynamic Smagorinsky \cite{smagorinsky1963general, lilly1983stratified, bou2005scale} and Anisotropic Minimum Dissipation \cite{rozema2015minimum, vreugdenhil2018large} support nonhydrostatic direct and large eddy simulations of meter-scale phenomena in periodic, closed, or open domains.
Nonhydrostatic simulations are typically run with a rigid lid, but an experimental time-implicit free surface formulation is also supported.
Hydrostatic regional to global ocean simulations with an implicit or split-explicit free surface formulation and $z$ or $z^\star$ vertical coordinate are supported on rectilinear, latitude-longitude, tripolar \cite{murray1996explicit}, and cubed sphere \cite{adcroft2004implementation} grids with second-order and WENO-based vector invariant schemes \cite{silvestri2024new}, the Gent--McWilliams parameterization \cite{gent1990isopycnal}, horizontal biharmonic diffusivity, and $Ri$-based, one-equation \cite{wagner2023catke} and two-equation vertical mixing schemes \cite{umlauf2005second}.
Linear, quadratic \cite{roquet2015defining}, and polynomial \cite{roquet2015accurate} equations of state are supported (with arbitrary gravitational direction for nonhydrostatic simulations) as well as traditional, non-traditional, spherical, and $\beta$-plane Coriolis forces.
Lagrangian particles, Stokes drift, and biogeochemistry are supported.

Oceananigans' programmable interface facilitates composition with other software, and a growing ecosystem of companion packages has rapidly formed around it --- developed largely by students, postdocs, and independent collaborators working outside the core team.
That this ecosystem emerged while Oceananigans was still maturing, and before comprehensive validation, suggests that the barrier to extending ocean modeling software in Julia is low enough to justify building on a young platform.
The companion package ``NumericalEarth'' \cite{NumericalEarth-zenodo} provides a coupled modeling framework for integrating Oceananigans ocean models together with sea ice simulations and either prescribed or prognostic atmospheres, using fluxes computed according to Monin--Obukhov similarity theory \cite<for example,>{edson2014exchange}.
NumericalEarth also implements utilities for automatic, script-based downloading of bathymetric data, ocean and sea ice physical and biogeochemical reanalysis products, and atmospheric reanalysis.
ClimaSeaIce \cite{ClimaSeaIce-zenodo} provides a sea ice modeling system that uses Oceananigans utilities and abstractions to implement simple sea ice and snow thermodynamics with a viscoplastic rheology.
ClimaOcean couples Oceananigans and ClimaSeaIce with the ClimaAtmos \cite{yatunin2026climate} and ClimaLand \cite{deck2025climaland} Julia packages for global Earth System simulations.

We conclude in section~\ref{sec:conclusions} by describing outstanding problems and questions, outlining ongoing development work, and anticipating the next major innovations in ocean modeling that will someday render the present work obsolete.

\section{Oceananigans, the library}
\label{sec:the-library}

Oceananigans is a \textit{library} of tools for building models by writing programs called ``scripts''.
This design departs from typical monolithic interfaces that ingest lists of flags and parameters from non-executable text files.
By blending mathematical symbols with verbose natural language names, Oceananigans syntax tries to enable evocative scripting that approaches the effectiveness of writing for communicating computational science.

\subsection{Hello, ocean}

The process of learning Oceananigans best begins by running simple simulations.
Our first example in listing~\ref{list:first-script} sets up, runs, and visualizes a simulation of two-dimensional turbulence.
The 21 lines of listing~\ref{list:first-script} illustrate one of Oceananigans' main achievements: a numerical experiment may be completely described by a single script.
Moreover, Oceananigans' interface is modular and concise, so that the complexity of an Oceananigans script is uniquely determined by the complexity of the problem being solved.
To execute the code in listing~\ref{list:first-script}, we need to copy it into a file (call this, for example, \code{hello\_ocean.jl}) and execute it by typing \code{include("hello\_ocean.jl")} in a Julia session.

\begin{jllisting}[float, caption={\pretolerance=100 A Julia script that uses Oceananigans and the Julia plotting library CairoMakie to set up, run, and visualize a simulation of two-dimensional turbulence on a Graphics Processing Unit (GPU). The initial velocity field, defined on lines~11-12, consists of random numbers uniformly-distributed between $-1$ and $1$. The vorticity $\zeta = \d_x v - \d_y u$ is defined on line~18. The solution is visualized in figure~\ref{fig:first-impression}.}, label={list:first-script}]
using Oceananigans, CUDA # using CUDA allows us to use an Nvidia GPU

# The third dimension is "flattened" to reduce the domain from three to two dimensions.
topology = (Periodic, Periodic, Flat)
architecture = GPU() # CPU() works just fine too for this small example.
x = y = (0, 2π)
grid = RectilinearGrid(architecture; size=(256, 256), x, y, topology)

model = NonhydrostaticModel(grid; advection=WENO(order=9))

ϵ(x, y) = 2rand() - 1 # Uniformly-distributed random numbers between [-1, 1).
set!(model, u=ϵ, v=ϵ)

simulation = Simulation(model; Δt=0.01, stop_time=10)
run!(simulation)

u, v, w = model.velocities
ζ = ∂x(v) - ∂y(u)

using CairoMakie
heatmap(ζ, colormap=:balance, axis=(; aspect=1))
\end{jllisting}

Oceananigans scripts are conventionally organized into four sections.
The first three define the grid, model, and simulation, and conclude with execution of the simulation.
The fourth section, often implemented separately for complex or expensive simulations, performs post-processing and analysis.
In listing~\ref{list:first-script}, the grid defined on lines 4--7 determines the problem geometry, grid spacing, and machine architecture.
To use a CPU instead of a GPU, one writes \code{CPU()} in place of \code{GPU()} on line~5: no other changes to the script are required. Note that while our examples use CUDA.jl to run on NVIDIA GPUs, Oceananigans also supports AMD and Intel GPUs (e.g. ALCF's Aurora) and, more experimentally, tensor processing units (TPUs) through Reactant cross-compilation \cite{moses2026adjoint}.

Lines 9--12 define the model, which solves the Navier--Stokes equations in two dimensions with a 9th-order Weighted, Essentially Non-Oscillatory (WENO) advection scheme (see~\ref{sec:finite-volume} and also \citeA{silvestri2024new} for more information about WENO).
The velocity components $u$ and $v$ are initialized with uniformly distributed random numbers within $[-1, 1)$.
The model definition can also encompass forcing, boundary conditions, and the specification of additional terms in the momentum and tracer equations such as Coriolis forces or turbulence closures.

Line~14 builds a simulation with a time-step $\Delta t = 0.01$ which will run until $t = 10$. (Time is non-dimensional in this example because the user has not specified physical units and because, by design, \code{NonhydrostaticModel} has no default parameters that could impose a unit system. Later examples illustrate how, with dimensional user input, variables like the time step and stop time can acquire user-specified dimensions.)
Lines~17-18 analyze the final state of the simulation by computing vorticity, illustrating Oceananigans' toolbox for composing discrete calculus and arithmetic operations.
The same tools may be used to define online diagnostics to be periodically computed and saved to disk while the simulation runs.
Line~21 concludes the numerical experiment with a visualization, which is shown in figure~\ref{fig:first-impression}.

\subsection{Incorporating user code}

\begin{figure}[htp]
\centering
\includegraphics[width=1.0\textwidth]{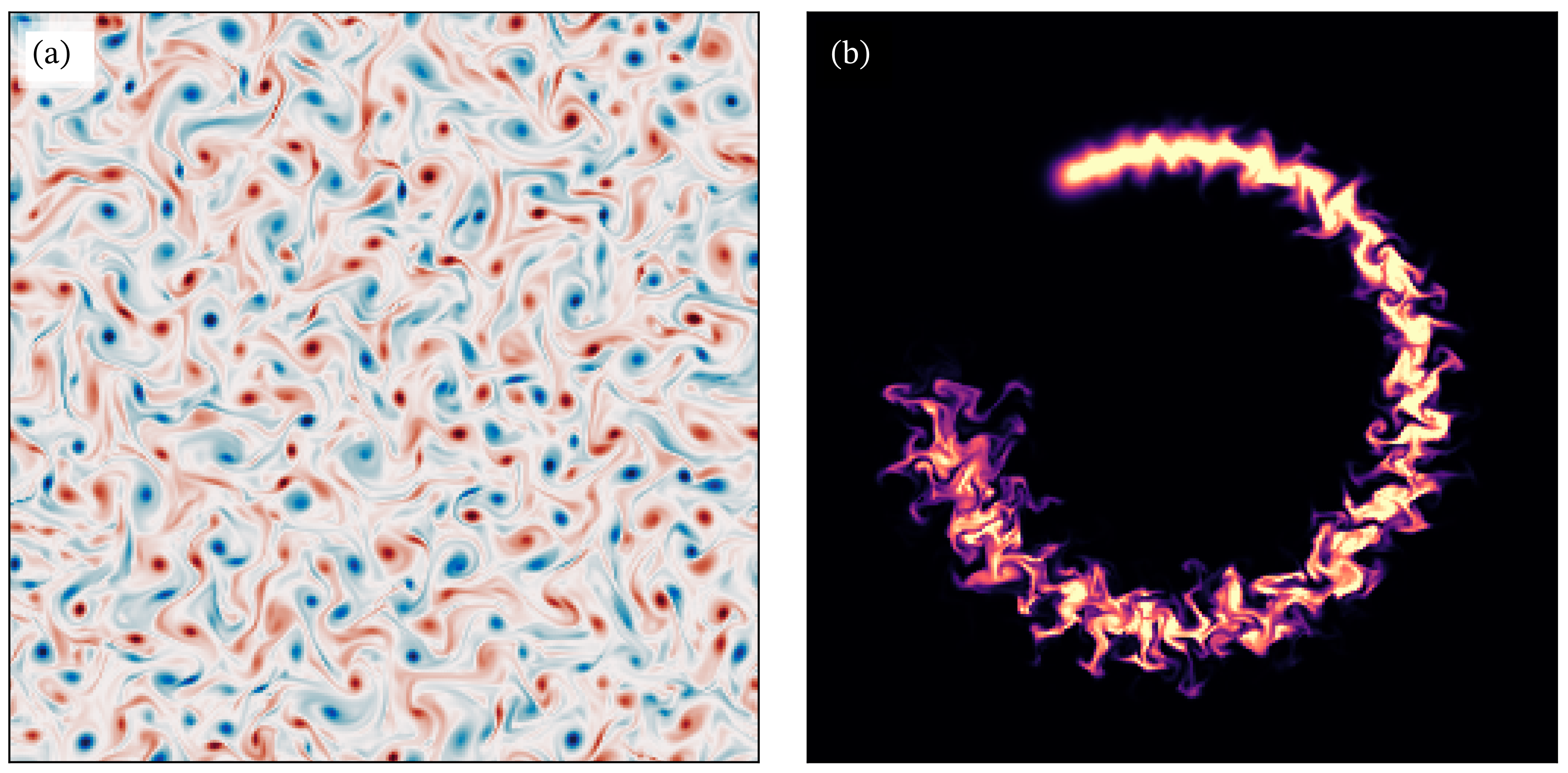}
\caption{\textit{(a)} Vorticity after $t=10$ and \textit{(b)} a passive tracer injected by a moving source at $t=2.5$ in a simulation of two-dimensional turbulence using an implicitly-dissipative advection scheme.
The figures are oriented so that $x$ varies with the horizontal axis and $y$ varies with the vertical axis (see listings~\ref{list:first-script} and~\ref{list:moving-source}).}
\label{fig:first-impression}
\end{figure}

With a programmable interface and aided by Julia's just-in-time compilation, user functions specifying domain geometry, forcing, boundary conditions, and initial conditions can be incorporated directly into models without modifying source code or using a separate programming environment.
To illustrate function-based forcing, we modify listing~\ref{list:first-script} with code that adds a passive tracer which is forced by a moving source that depends on $x, y, t$.
A visualization of the vorticity and tracer field generated by listings~\ref{list:first-script} and~\ref{list:moving-source} is shown in figure~\ref{fig:first-impression}.

\begin{jllisting}[float, caption={Implementation of a moving source of passive tracer with a function in a two-dimensional turbulence simulation. These lines of code replace the model definition on line~9 in listing~\ref{list:first-script}. Julia supports Unicode variable names, including Greek letters.}, label={list:moving-source}]
function circling_source(x, y, t)
    δ, ω, r = 0.1, 2π/3, 2
    dx = x + r * cos(ω * t)
    dy = y + r * sin(ω * t)
    return exp(-(dx^2 + dy^2) / 2δ^2)
end

forcing = (; c = circling_source)
model = NonhydrostaticModel(; grid, advection=WENO(order=9), tracers=:c, forcing)
\end{jllisting}

General tasks may be inserted into the time stepping loop by modifying the simulation.
This supports things as mundane as printing a summary of the current model status or writing output, to more exotic tasks like nudging state variables or updating a diffusion coefficient based on an externally-implemented model.

Writing simulation output to disk is a well-known bottleneck in high-performance computing \cite<see, for example,>{lofstead2010managing}.
Similar to approaches based on dedicated I/O servers \cite{dennis2012computational}, Oceananigans utilizes host CPUs to perform I/O concurrently with continued GPU computation. As with dedicated I/O servers, overlapping disk writes with computation can reduce output overhead, but does not eliminate data-transfer costs or limits imposed by sustained storage throughput.

\subsection{Abstractions for arithmetic and discrete calculus}

Abstractions representing unary, binary, and calculus operators produce a system for building ``lazy'' expression trees that are evaluated only upon request (for example, if their output should be periodically saved to disk during a simulation).
Example calculations representing vorticity, $\zeta = \d_x v - \d_y u$, speed $s = \sqrt{u^2 + v^2}$, and the $x$-integral of enstrophy $Z = \int_0^{2\pi} \zeta^2 \di x$ are shown in listing~\ref{list:expressions}, alongside an example that uses the same staggered-grid stencil operators found in Oceananigans source code to compute the gradient Richardson number in a user script via \code{KernelFunctionOperation}. Oceananigans and Breeze \cite{Breeze-zenodo} provide built-in diagnostics, while Oceanostics \cite{oceanostics} provides additional reusable diagnostics built on these abstractions, so users need not implement each diagnostic in their simulation scripts.

\begin{jllisting}[caption={Illustration of abstractions for expression trees, reductions, and user-defined stencil computations called ``\code{AbstractOperations}'' in Oceananigans. \code{AbstractOperations} are ``lazy'', in the sense that they \textit{represent} potential computations but do not instantiate the computation directly. The first half shows simple expression trees for vorticity, speed, and enstrophy. The second half computes the gradient Richardson number using the same staggered-grid interpolation and differentiation operators found in Oceananigans source code; \code{KernelFunctionOperation} provides the bridge that wraps these source-code-level operators into the \code{AbstractOperation} interface for use in user scripts.}, label={list:expressions}]
u, v, w = model.velocities

# "Lazy" expression trees and reductions representing computations:
ζ = ∂x(v) - ∂y(u)
s = √(u^2 + v^2)
Z = Integral(ζ^2, dims=1)

# Building and computing a Field that instantiates an AbstractOperation:
ζ_field = Field(∂x(v) - ∂y(u))

# KernelFunctionOperations: user-defined stencil computations
@inline ϕ²(i, j, k, grid, ϕ, args...) = ϕ(i, j, k, grid, args...)^2

@inline function Ri(i, j, k, grid, velocities, buoyancy, tracers)
    ∂z_u² = ℑxᶜᵃᵃ(i, j, k, grid, ϕ², ∂zᶠᶜᶠ, velocities.u)
    ∂z_v² = ℑyᵃᶜᵃ(i, j, k, grid, ϕ², ∂zᶜᶠᶠ, velocities.v)
    S² = ∂z_u² + ∂z_v²
    N² = ∂z_b(i, j, k, grid, buoyancy, tracers)
    Ri = N² / S²
    return ifelse(S² == 0, zero(grid), Ri)
end

Ri = KernelFunctionOperation{Center, Center, Face}(Ri, grid,
         model.velocities, model.buoyancy, model.tracers)

# Save diagnostics periodically after computing online on GPU
output_writer = JLD2Writer(model, (; ζ, Ri), schedule=TimeInterval(1hour))
\end{jllisting}

The second half of listing~\ref{list:expressions} illustrates the staggered-grid interpolation operators $\mathcal{I}_x^{caa}$, $\mathcal{I}_y^{aca}$ and differentiation operators $\partial_z^{fcf}$, $\partial_z^{cff}$ that encode the finite volume staggering of the C-grid.
These are the same operators used throughout the Oceananigans source code to build numerical algorithms --- for example, the Pacanowski--Philander vertical mixing scheme \cite{pacanowski1981parameterization} is implemented with identical stencil calls.
\code{KernelFunctionOperation} provides the bridge that exposes these source-code-level operators to user scripts: wrapping a hand-written kernel function into the \code{AbstractOperation} interface so that the resulting $Ri$ object can be saved to disk, reduced, or composed with further operations, just like the simple expression trees above.
The example computes the gradient Richardson number $Ri = N^2 / S^2$ using the same numerical operators as internal parameterizations. Sharing these operators between model internals, user-defined diagnostics, and companion packages allows numerical code to be reused across applications without duplication.
(Note, a more stable $Ri$ stencil is described by \citeA{griffies2025cm4x}.)

\section{Governing equations, parameterizations, and illustrative examples}

\label{sec:equations-parameterizations}

Oceananigans implements two ``models'' for ocean-flavored fluid dynamics: the \code{HydrostaticFreeSurfaceModel}, and the \code{NonhydrostaticModel}.
Each represents a template for equations that govern the evolution of momentum and tracers.
Both models are incompressible and make the Boussinesq approximation, which means that the density of the modeled fluid is decomposed into a constant reference $\rho_0$ and a small dynamic perturbation $\rho'$,
\beq \label{boussinesq}
\rho(\bx, t) = \rho_0 + \rho'(\bx, t) \com
\qquad \text{where} \qquad 
|\rho'| \ll \rho_0 \com
\eeq
and $\bx = (x, y, z)$ is position and $t$ is time.

The relative smallness of $\rho'$ reduces conservation of mass to a statement of incompressibility called the continuity equation,
\beq \label{continuity}
\bnabla \bcdot \bu = 0 \com
\eeq
where
\beq
\bu \defn u \bxh + v \byh + w \bzh \com
\eeq
is the three-dimensional velocity field.
Within the Boussinesq approximation, the momentum $\rho_0 \bu$ varies only with the velocity $\bu$.
The effect of density variations is encapsulated by a buoyant acceleration,
\beq \label{buoyancy}
b \defn - \frac{g \rho'}{\rho_0} \com
\eeq
where $g$ is gravitational acceleration.
The ``buoyancy'' $b$ acts against the direction of gravity.

For gravity directed along $-\bzh$, the total pressure $P$, relative to a constant reference pressure, is decomposed into
\beq \label{pressure}
P = -\rho_0 g z + \rho_0 p(\bx, t) \com
\eeq
where $z$ is height, $-\rho_0 g z$ is the static contribution to pressure that opposes the gravitational force associated with the reference density $\rho_0$, and $\rho_0 p$ represents the dynamic anomaly.
$p$ is called the kinematic pressure.

\subsection{The \code{NonhydrostaticModel}}

The \code{NonhydrostaticModel} represents the Boussinesq equations formulated \textit{without} making the hydrostatic approximation typical to general circulation models.
The \code{NonhydrostaticModel} has a three-dimensional prognostic velocity field.

\subsubsection{The \code{NonhydrostaticModel} momentum equation}

The \code{NonhydrostaticModel}'s momentum equation incorporates advection by a background velocity field, Coriolis forces, surface wave effects via the Craik-Leibovich asymptotic model \cite{craik1976rational, huang1979surface}, a buoyancy term allowed to be a nonlinear function of tracers and depth, a stress divergence derived from molecular friction or a turbulence closure, and a user-defined forcing term.
Using the Boussinesq approximation in \eqref{boussinesq} and the pressure decomposition in \eqref{pressure}, the generic form of \code{NonhydrostaticModel}'s momentum equation is
\beq \label{nonhydrotatic-momentum-equations}
\begin{split}
\d_t \bu &=
- \bnabla p
\, - \underbrace{\strut \left ( \bu \bcdot \bnabla \right ) \bu - \left ( \bubg \bcdot \bnabla \right ) \bu - \left ( \bu \bcdot \bnabla \right ) \bubg}_{\text{advection}}
\quad - \quad \underbrace{\strut \b{f} \times \bu}_{\text{Coriolis}} \\[2ex]
& + \, \underbrace{\strut \left ( \bnabla \times \buS \right ) \times \bu + \d_t \buS}_{\text{Stokes forcing}}
\quad - \underbrace{\strut b \, \b{\hat g}}_{\text{buoyancy}}
- \quad \underbrace{\strut \bnabla \bcdot \b{\tau}}_{\text{closure}}
\quad + \underbrace{\strut \b{F}_u}_{\text{forcing}} \com
\end{split}
\eeq
where $\bubg$ is a prescribed and divergence-free ``background'' velocity field, $p$ is the kinematic pressure, $\b{f}$ is the background vorticity associated with a rotating frame of reference, $\buS$ is the Stokes drift profile associated with a prescribed surface wave field, $b$ is buoyancy, $\b{\hat g}$ is the gravitational unit vector (usually pointing downwards, that is, $\b{\hat g} = - \bzh$), $\b{\tau}$ is the stress tensor associated with molecular viscous or subgrid turbulent momentum transport, and $\b{F}_u$ is a body force.

To integrate equation~\eqref{nonhydrotatic-momentum-equations} while enforcing \eqref{continuity}, we use a pressure-correction method that requires solving a three-dimensional Poisson equation to find $p$, which can be derived by taking the divergence of \eqref{nonhydrotatic-momentum-equations}.
This Poisson equation is often a computational bottleneck in curvilinear or irregular domains, and its elimination is the main motivation for making the hydrostatic approximation when formulating the \code{HydrostaticFreeSurfaceModel}, as described in section~\ref{sec:hydrostatic-model-equations}.
For rectilinear grids, we solve the Poisson equation using a direct FFT-based or mixed FFT-tridiagonal solver \cite{schumann1988fast}, providing substantial acceleration over MITgcm's conjugate gradient pressure solver \cite{marshall1997finite}.
The FFT-based pressure solver uses the distributed transpose algorithm described in \ref{app:parallelization} and figure~\ref{fig:transpose}, and has been used in simulations on 1,024 GPUs \cite{peng2026submesoscale}.
Detailed scaling results will be presented in a forthcoming study of FFT-based solvers for complex domains.
In irregular domains, we either use a masking method that permits an approximate solution of the pressure Poisson equation with the FFT-based method, or an iterative conjugate gradient solver that leverages the FFT-based solver as a preconditioner.
The pressure correction scheme is described further in~\autoref{fractional-step-method}.

Advection in the \code{NonhydrostaticModel} may be formulated in ``flux form'',
\beq \label{flux-form-advection}
\text{advection} = u_j \d_j u_i + {\ubg}_j \d_j u_i + u_j \d_j {\ubg}_i = \d_j \Big [ \big (u_j + {\ubg}_j \big ) u_i + u_j {\ubg}_i \Big ] \com
\eeq
where we have used the continuity equation~\eqref{continuity}. We use indicial notation; for example, the $i$-th component of the advection term is $\left [ \left ( \bu \bcdot \bnabla \right ) \bu \right]_i = u_j \d_j u_i$.
(See the text surrounding equations~\eqref{hydrostatic-flux-advection}--\eqref{hydrostatic-weno-vi-advection} for a discussion of advection term formulation in the \code{HydrostaticFreeSurfaceModel}.)

The formulation of the Stokes drift terms in~\eqref{nonhydrotatic-momentum-equations} means that $\bu$ is the Lagrangian-mean velocity when Stokes drift effects are included \cite<see, for example,>{wagner2021near}.
With a Lagrangian-mean formulation, equations~\eqref{continuity} and \eqref{nonhydrotatic-momentum-equations} are consistent only when $\buS$ is non-divergent --- or equivalently, when $\buS$ is obtained by projecting the divergence out of the usual Stokes drift \cite{vanneste2022stokes}. A consistent two-way coupling between surface waves and currents that conserves momentum and energy is developed by \citeA{vanneste2026consistent}.
As discussed by \citeA{wagner2021near}, the Lagrangian-mean formulation of \eqref{nonhydrotatic-momentum-equations} means that closures for LES strictly dissipate resolved kinetic energy, avoiding the inconsistency between resolved and subgrid fluxes affecting typical LES formulated in terms of the Eulerian-mean velocity \cite<see also>{pearson2018turbulence, wagner2025phenomenology, li2026stabilizing}.

The labeled terms in \eqref{nonhydrotatic-momentum-equations} are controlled by arguments to \code{NonhydrostaticModel} invoked in both of listings~\ref{list:first-script} and \ref{list:moving-source}.
For example, ``advection'' chooses a numerical scheme to approximate the advection term in~\eqref{nonhydrotatic-momentum-equations} and \eqref{flux-form-advection}.
As another example, we consider configuring the closure term in~\eqref{nonhydrotatic-momentum-equations} to represent \textit{(i)} molecular diffusion by a constant-coefficient Laplacian \code{ScalarDiffusivity}, \textit{(ii)} turbulent stresses approximated by the \code{SmagorinskyLilly} eddy viscosity model \cite{smagorinsky1963general, lilly1983stratified} for large eddy simulation, or \textit{(iii)} omitting it entirely and relying on WENO advection schemes for dissipation and stability \cite{pressel2017numerics, wagner2023catke}.
In these three cases, the closure flux divergence $\bnabla \bcdot \b{\tau} = \d_m \tau_{nm}$ in indicial notation becomes
\beq
- \d_m \tau_{n m} = \left \{
\begin{matrix}
\d_m \left ( \nu \d_m u_n \right ) & \qquad & \text{(\code{ScalarDiffusivity})} \\
0 & \qquad & \text{(nothing)} \\
\d_m \Big ( 2 \underbrace{\mystrut{1ex} C_s \Delta^2 |\Sigma|}_{\nu_e} \Sigma_{nm} \Big ) & \qquad & \text{(\code{SmagorinskyLilly})}
\end{matrix} \right .
\eeq
where $\nu$ is the Laplacian viscosity, $\Sigma_{nm} = \tfrac{1}{2} (\d_m u_n + \d_n u_m)$ is the strain rate tensor, $|\Sigma|$ is the magnitude of the strain rate tensor, $C_s$ is the \code{SmagorinskyLilly} model constant, $\Delta$~scales with the local grid spacing, and $\nu_e$ is the eddy viscosity.

\code{ScalarDiffusivity} diffusion coefficients may also vary in time- and space or depend on model fields.
Other closure options for \code{NonhydrostaticModel} include fourth-order \code{ScalarBiharmonicDiffusivity}, various Lagrangian-averaged or directionally-averaged flavors of \code{DynamicSmagorinsky} \cite{bou2005scale}, and the \code{AnisotropicMinimumDissipation} turbulence closure \cite{rozema2015minimum, abkar2016minimum, vreugdenhil2018large, wagner2021near} for large eddy simulations.
\code{SmagorinskyLilly} and \code{DynamicSmagorinsky} use Laplacian, rather than biharmonic, eddy viscosity.
Vertically-implicit time-discretization is supported with all scalar-type diffusivities, including \code{ScalarDiffusivity} and LES closures.

We note that LES may be conducted solely with WENO advection schemes that dissipate grid-scale kinetic energy undergoing a forward cascade from large to small scales.
However, no reliable method has yet been developed to diagnose the dissipation of kinetic energy in such ``implicit LES'' simulations, so, at the time of writing, explicit closures must be included if such diagnostics are required.

We illustrate how \code{NonhydrostaticModel} may be used to model fluid motion on the smallest scales in listing~\ref{list:flow-around-cylinder}, which implements a direct numerical simulation of uniform flow past a cylinder with no-slip boundary conditions.
The goal of direct numerical simulation is to resolve all scales of motion down to the viscous dissipation scale.
For this purpose, we use molecular viscosity with a fourth-order diffusion stencil, \code{ScalarDiffusivity} with \code{stencil\_order=4}, and fourth-order centered advection, \code{Centered(order=4)}. These orders describe the individual interior operators; the stencils revert to second order near walls and immersed boundaries, so we do not claim global fourth-order accuracy.
Lines~6--7 of listing~\ref{list:flow-around-cylinder} embed a cylindrical mask in a \code{RectilinearGrid} using a \code{GridFittedBoundary}, which generalizes to arbitrary three-dimensional shapes.
The no-slip condition is specified on lines~11--12 by invoking \code{ValueBoundaryCondition} (a synonym for ``Dirichlet'' boundary condition): \code{ValueBoundaryCondition} enforces a near-wall tangential velocity gradient, which in turn produces a viscous flux associated with the \code{ScalarDiffusivity} viscosity \cite<resolving to the same formulation proposed in a seminal paper by>{harlow1965numerical}.
Other choices include \code{GradientBoundaryCondition} (Neumann), \code{MixedBoundaryCondition} (Robin condition on a linear combination of the scalar value and gradient), \code{FluxBoundaryCondition} (direct imposition of fluxes), \code{OpenBoundaryCondition} (for non-trivial boundary-normal velocity fields), as well as a helper function \code{BulkDrag} (for quadratic drag fluxes, which are a kind of \code{FluxBoundaryCondition}).

\begin{jllisting}[float, caption={\pretolerance=100 Direct numerical simulation of flow past a cylinder at various Reynolds numbers $Re$. The domain is periodic in $x$ and a sponge layer on the right side of the domain relaxes the solution to $\bu = u_\infty \bxh$ with $u_\infty = 1$. The experiment can be converted to a large eddy simulation (thereby sending $Re \to \infty$) by replacing the no-slip boundary conditions with an appropriate drag model and either \textit{(i)} using an appropriate turbulence closure or \textit{(ii)} using the \code{WENO(order=9)} advection scheme with no turbulence closure. Visualizations of the DNS and LES cases are shown in figure~\ref{fig:flow-around-cylinder}.}, label={list:flow-around-cylinder}]
r, U, Re, Ny = 1/2, 1, 1000, 2048

grid = RectilinearGrid(GPU(), size=(2Ny, Ny), x=(-3, 21), y=(-6, 6),
    topology=(Periodic, Bounded, Flat))

cylinder(x, y) = (x^2 + y^2) ≤ r^2
grid = ImmersedBoundaryGrid(grid, GridFittedBoundary(cylinder))

closure = ScalarDiffusivity(ν=1/Re, stencil_order=4)

no_slip = FieldBoundaryConditions(immersed=ValueBoundaryCondition(0))
boundary_conditions = (u=no_slip, v=no_slip)

# Implement a sponge layer on the right side of the domain that
# relaxes v → 0 and u → U over a region of thickness δ
@inline mask(x, y, δ=3, x₀=21) = max(zero(x), (x - x₀ + δ) / δ)
Fu = Relaxation(target=U; mask, rate=1)
Fv = Relaxation(target=0; mask, rate=1)

model = NonhydrostaticModel(; grid, closure, advection=Centered(order=4),
    boundary_conditions, forcing=(u=Fu, v=Fv))
\end{jllisting}

\begin{figure}[htp]
\centering
\includegraphics[width=\textwidth]{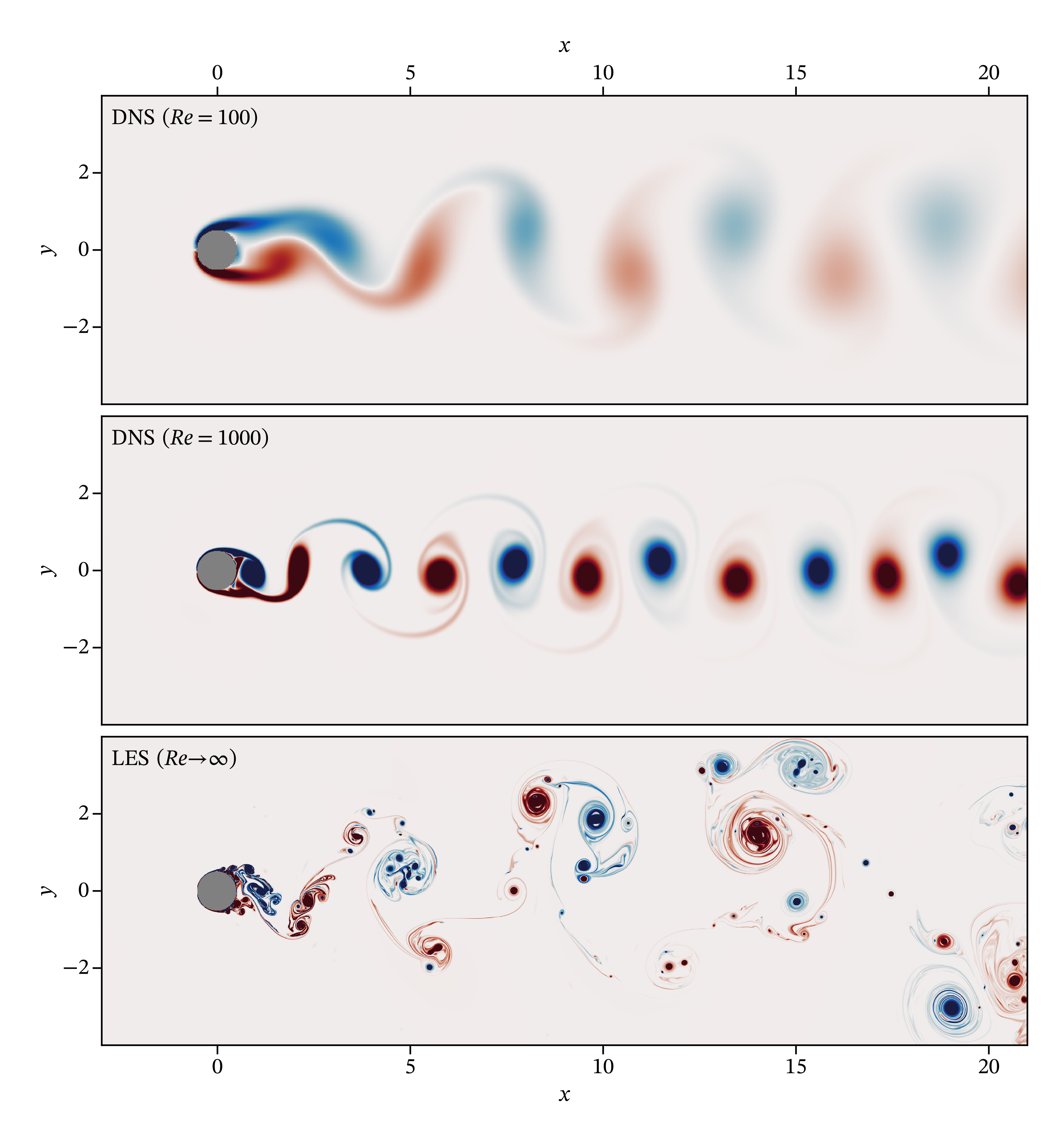}
\caption{Vorticity snapshots in simulations of flow around a cylinder. The top two panels show vorticity in direct numerical simulations (DNS) that use a molecular \code{ScalarDiffusivity} closure with a fourth-order diffusion stencil and \code{Centered(order=4)} advection with Reynolds numbers (Re) of 100 (top panel) and 1000 (middle panel). The bottom panel shows a large eddy simulation (LES) with no closure and a \code{WENO(order=9)} advection scheme. Note that the sponge layer could be replaced with a \code{PerturbationAdvection} open boundary condition if desired. The figures are oriented so that $x$ varies in the horizontal direction and $y$ varies in the vertical direction.}
\label{fig:flow-around-cylinder}
\end{figure}

Results obtained with listing~\ref{list:flow-around-cylinder} for $Re=100$, $Re=1000$, and a modified version of listing~\ref{list:flow-around-cylinder} for large eddy simulation ($Re \to \infty)$ are visualized in figure~\ref{fig:flow-around-cylinder}.
To adapt listing~\ref{list:flow-around-cylinder} for LES, the closure is eliminated in favor of a 9th-order WENO advection scheme, and the no-slip boundary condition is replaced with a quadratic drag boundary condition with a drag coefficient estimated from similarity theory using a constant estimated roughness length.

\subsubsection{The \code{NonhydrostaticModel} tracer conservation equation}

The buoyancy term in \eqref{nonhydrotatic-momentum-equations} requires tracers, and can be formulated to use buoyancy itself as a tracer, or to depend on temperature $T$ and salinity $S$.
For seawater, a 54-term polynomial approximation \code{TEOS10EquationOfState} \cite{mcdougall2011getting, roquet2015accurate} is implemented in the auxiliary package SeawaterPolynomials, along with quadratic approximations to TEOS-10 \cite{roquet2015defining} and a \code{LinearEquationOfState}.
All tracers --- either ``active'' tracers required to compute the buoyancy term, as well as additional user-defined passive tracers --- obey the tracer conservation equation
\beq \label{nonhydrostatic-tracer-equation}
\d_t c =
\underbrace{\strut - \bnabla \bcdot \left [ \left ( \bu + \bubg + \bubgc \right ) c + \bu \cbg \right ] + c \bnabla \bcdot \bubgc}_{\text{advection}}
\, - \, \underbrace{\strut \bnabla \bcdot \b{J}_c}_{\text{closure}}
\, + \underbrace{\strut S_c}_{\text{biogeochemistry}}
+ \underbrace{\strut F_c}_{\text{forcing}} \com
\eeq
where $c$ represents any tracer, $\cbg$ represents a prescribed background tracer concentration for $c$, $\bubgc$ is the velocity associated with biogeochemical processes (and which, in general, can have non-zero divergence), $\b{J}_c$ is a tracer flux associated with molecular diffusion or subgrid turbulence, $S_c$ is a source or sink term associated with biogeochemical processes, such as chemical reactions and biological growth and mortality, \cite<provided, for example, by external packages like OceanBioME; >{strong2023oceanbiome}, and $F_c$ is a user-defined source or sink.
The formulation of tracer advection in~\eqref{nonhydrostatic-tracer-equation} leverages $\bnabla \bcdot \bu = 0$ and $\bnabla \bcdot \bubg = 0$.

\begin{figure}[htp]
\centering
\includegraphics[width=1\textwidth]{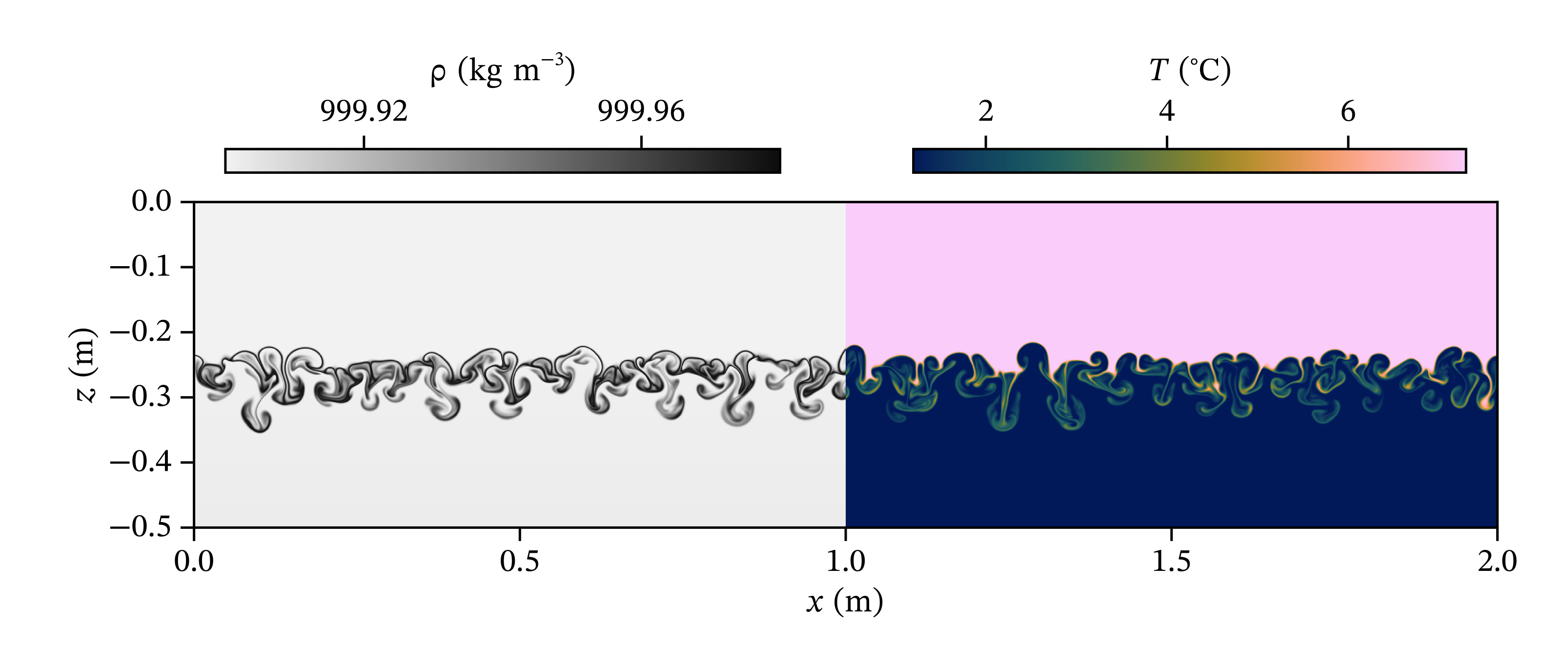}
\caption{Density and temperature at $t=1$ minute in a direct numerical simulation of cabbeling in freshwater. Note that both fields span from $x=0$ to $x=2$ meters; only the left half of the density field and the right half of the temperature field are shown.}
\label{fig:cabbeling}
\end{figure}

A simulation with a passive tracer having a user-defined source term is illustrated by listing~\ref{list:moving-source} and figure~\ref{fig:first-impression}.
For a second example, we consider freshwater cabbeling \cite<see, for example,>{Bisits2025cabbeling}.
Cabbeling occurs when two water masses of similar density mix to form a new water mass which, due to the nonlinearity of the equation of state, is denser than either of its constituents.
Freshwater, for example, is densest at 4 degrees Celsius, while 1- and 7.55-degree water are lighter with roughly the same density.
We implement a direct numerical simulation (DNS) in which 7.55-degree water overlies 1-degree water, using the \code{TEOS10EquationOfState} provided by the auxiliary package SeawaterPolynomials.
The script is shown in listing~\ref{list:cabbeling}.
The resulting density and temperature fields after 1 minute of simulation are shown in figure~\ref{fig:cabbeling}.
Note that the \code{TEOS10EquationOfState} typically depends on both temperature and salinity tracers, but listing~\ref{list:cabbeling} specifies a constant salinity $S=0$ and thus avoids allocating memory for or simulating salinity directly.
Although fourth-order advection and diffusion operators are available, other components of the finite-volume formulation remain second order; these operators alone do not provide the accuracy of pseudospectral formulations for DNS \cite{lecoanet2016validated}.
(This contrasts with Oceananigans' capabilities for large eddy simulation, where simulation error is dominated by the representation of grid-scale dissipation rather than the formal accuracy of the discretization.)

\begin{jllisting}[float, caption={Direct numerical simulation of convective turbulence driven by cabbeling between 1- and 7.55-degree freshwater. $\nu$ denotes viscosity and $\kappa$ denotes the tracer diffusivity. The diffusivity may also be set independently for each tracer.}, label={list:cabbeling}]
grid = RectilinearGrid(GPU(), topology = (Bounded, Flat, Bounded),
    size = (4096, 1024), x = (0, 2), z = (-0.5, 0))

closure = ScalarDiffusivity(ν=1.15e-6, κ=1e-7)

using SeawaterPolynomials: TEOS10EquationOfState
equation_of_state = TEOS10EquationOfState(reference_density=1000)

buoyancy = SeawaterBuoyancy(gravitational_acceleration = 9.81,
    constant_salinity = 0, # set S=0 and simulate T only
    equation_of_state)

model = NonhydrostaticModel(; grid, buoyancy, closure, tracers=:T,
    advection=Centered(order=2), timestepper=:RungeKutta3)

Tᵢ(x, z) = z > -0.25 ? 7.55 : 1
Ξᵢ(x, z) = 1e-2 * randn()
set!(model, T=Tᵢ, u=Ξᵢ, v=Ξᵢ, w=Ξᵢ)
\end{jllisting}

We next consider a large eddy simulation of the Eady problem \cite{eady1949long}.
In the Eady problem, perturbations evolve around a basic state with constant shear $\Lambda$ in thermal wind balance with a constant meridional buoyancy gradient $f \Lambda$, such that
\beq
u = \underbrace{\mystrut{0.5ex} \Lambda z}_{\defn U} \, + \, u' \com
\qquad \text{and} \qquad
b = \underbrace{\mystrut{0.5ex} - f \Lambda y}_{\defn B} \, + \, b' \per
\eeq
We use Oceananigans' \code{BackgroundFields} to simulate the nonlinear evolution of $(u', v, w)$ and $b'$ expanded around $U$ and $B$ in a doubly-periodic domain.
We impose an initially stable density stratification with $b' = N^2 z$ and $N^2 = 10^{-6} \, \mathrm{s^{-2}}$ superposed with random noise.
The Richardson number of the initial condition is $Ri = N^2 / (\d_z U)^2 = N^2 / \Lambda^2$; we choose mean shear~$\Lambda$ so that $Ri = 1$, which guarantees the basic state is unstable to baroclinic instability but stable to symmetric and Kelvin-Helmholtz instability \cite{stone1971baroclinic}.
A portion of the script is shown in listing~\ref{list:eady-les}.

\begin{jllisting}[float, caption={Large eddy simulation of the Eady problem expanded around the background geostrophic shear with $Ri = 1$.}, label={list:eady-les}]
grid = RectilinearGrid(GPU(); size = (1024, 1024, 64),
    x = (0, 4096), y = (0, 4096), z = (0, 128),
    topology = (Periodic, Periodic, Bounded))

f, N², Ri = 1e-4, 1e-6, 1
parameters = (f=f, Λ=sqrt(N²/Ri)) # U = Λz, so Ri = N² / ∂z(U)^2 = N² / Λ^2 and Λ = N / √Ri.

@inline U(x, y, z, t, p) = + p.Λ * z
@inline B(x, y, z, t, p) = - p.f * p.Λ * y

background_fields = (u = BackgroundField(U; parameters),
    b = BackgroundField(B; parameters))

model = NonhydrostaticModel(; grid, background_fields,
    advection = WENO(order=9), coriolis = FPlane(; f),
    tracers = :b, buoyancy = BuoyancyTracer())

Δz = minimum_zspacing(grid)
bᵢ(x, y, z) = N² * z + 1e-2 * N² * Δz * (2rand() - 1)
set!(model, b=bᵢ)
\end{jllisting}

\begin{figure}[htp]
\centering
\includegraphics[width=1\textwidth]{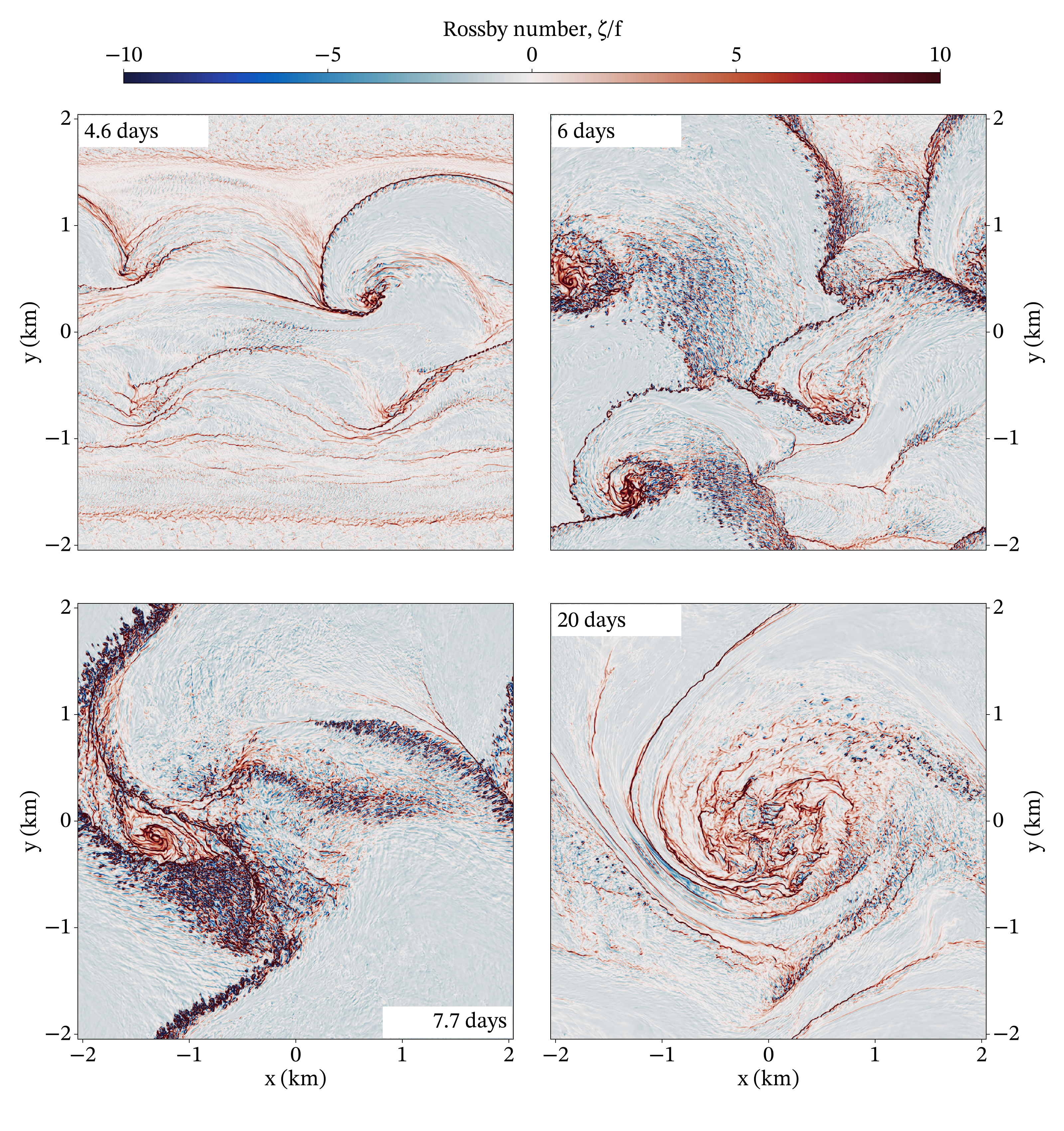}
\caption{Surface vertical vorticity in a large eddy simulation of the Eady problem with $Ri = 1$ initially, after $t = 4.6$, 6, 7.7, and 20 days. The grid spacing is $4 \times 4 \times 2$ meters in $x, y, z$. Part of the script that produces this simulation is shown in listing~\ref{list:eady-les}.}
\label{fig:eady-les}
\end{figure}

Our Eady simulation uses fully-turbulence-resolving resolution with 4 meter horizontal spacing and 2 meter vertical spacing in a $4 \, \mathrm{km} \times 4 \, \mathrm{km} \times 128 \, \mathrm{m}$ domain, $1024 \times 1024 \times 64$ grid points, and simulates 30~days on a single Nvidia H100 GPU.
On this grid and with 9th-order WENO advection, the model achieves around 19.2 simulated days per day (SDPD), corresponding to 0.84 wall-time-seconds per time-step and a time step size $\Delta t = 16.2$ s computed adaptively to target a Courant-Friedrichs-Lewy (CFL) number of $0.7$, after the transition to turbulence.
Four snapshots of vertical vorticity normalized by~$f$ (the Rossby number) are shown in figure~\ref{fig:eady-les}, illustrating the growth of kilometer-scale vortex motions amid bursts of meter-scale three-dimensional turbulence that develop along thin filaments of vertical vorticity and vertical shear.
This configuration captures a competition between baroclinic instability, which acts to ``restratify'' or strengthen boundary layer stratification, and three-dimensional turbulent mixing driven by a forward cascade from kilometer-scale motions \cite{molemaker2010balanced, dong2024submesoscales}.
Shear and convection associated with atmospheric storms provide an additional important source of ocean turbulent mixing interacting with kilometer-scale baroclinic instability \cite{boccaletti2007mixed, callies2018baroclinic}; these additional processes can be simulated by adding surface fluxes to our Eady configuration.

Finally, we illustrate Oceananigans' capabilities for realistic, three-dimensional large eddy simulations in complex geometries by simulating temperature- and salinity-stratified tidal flow past a headland, reminiscent of an extensively observed and modeled flow past Three Tree Point in Puget Sound in the Pacific Northwest of the United States \cite{pawlak2003observations, warner2014dynamics}. 
The bathymetry involves a sloping wedge that juts from a square-sided channel, such that
\beq
z_b(x, y) = -H \left ( 1 + \frac{y + |x|}{\delta} \right ) \com
\eeq
where $\delta = L/2$ represents the scale of the bathymetry, $L$ is the half-channel width in $y$ (the total width is $2L$), and $H = 256 \, \mathrm{m}$ is the depth of the channel, and $z=z_b(x, y)$ is the height of the bottom.
The flow is driven by a tidally-oscillating boundary velocity
\beq
U(t) = U_2 \sin \left ( \frac{2 \pi t}{\tau_2} \right )
\eeq
imposed at the east and west boundaries. Here, $\tau_2 = 12.421 \, \mathrm{hours}$ is the period of the semi-diurnal lunar tide, and $U_2 = 0.1 \, \mathrm{m \, s^{-1}}$ is the tidal velocity amplitude used in this idealized simulation.
The initial temperature and salinity are
\beq
T \, |_{t=0} = 12 + 4 \frac{z}{H} \, \mathrm{{}^\circ C} \com
\qquad \text{and} \qquad
S \, |_{t=0} = 32 \, \mathrm{g \, kg^{-1}} \per
\eeq
A portion of the script that implements this simulation is shown in listing~\ref{list:headland-les}.

\begin{jllisting}[float, caption={Large eddy simulation of flow past a headland reminiscent of Three Tree Point in the Pacific Northwest \protect\cite<see>{pawlak2003observations, warner2014dynamics}.}, label={list:headland-les}]
H, L = 256meters, 1024meters
δ = L / 2
x, y, z = (-3L, 3L), (-L, L), (-H, 0)
Nz = 64

grid = RectilinearGrid(GPU(); size=(6Nz, 2Nz, Nz), halo=(6, 6, 6),
    x, y, z, topology=(Bounded, Bounded, Bounded))

wedge(x, y) = -H *(1 + (y + abs(x)) / δ)
grid = ImmersedBoundaryGrid(grid, GridFittedBottom(wedge))

τ₂ = 12.421hours
U₂ = 0.1 # m/s

@inline U(x, y, z, t, p) = p.U₂ * sin(2π * t / p.τ₂)
@inline U(y, z, t, p) = U(zero(y), y, z, t, p)

open_bc = OpenBoundaryCondition(U; parameters=(; U₂, τ₂),
    scheme=PerturbationAdvection(inflow_timescale=2minutes, outflow_timescale=2minutes))

u_bcs = FieldBoundaryConditions(east = open_bc, west = open_bc)

@inline ambient_temperature(x, z, t, H) = 12 + 4z/H
@inline ambient_temperature(x, y, z, t, H) = ambient_temperature(x, z, t, H)
ambient_temperature_bc=ValueBoundaryCondition(ambient_temperature; parameters=H)
T_bcs=FieldBoundaryConditions(east=ambient_temperature_bc, west=ambient_temperature_bc)

ambient_salinity_bc = ValueBoundaryCondition(32)
S_bcs=FieldBoundaryConditions(east=ambient_salinity_bc, west=ambient_salinity_bc)

model = NonhydrostaticModel(; grid, tracers = (:T, :S),
    buoyancy=SeawaterBuoyancy(equation_of_state=TEOS10EquationOfState()),
    advection=WENO(order=9), coriolis=FPlane(latitude=47.5),
    boundary_conditions=(; T=T_bcs, u=u_bcs, S=S_bcs))

Tᵢ(x, y, z) = ambient_temperature(x, y, z, 0, H)
set!(model, T=Tᵢ, S=32)
\end{jllisting}

\begin{figure}[htp]
\centering
\includegraphics[width=1\textwidth]{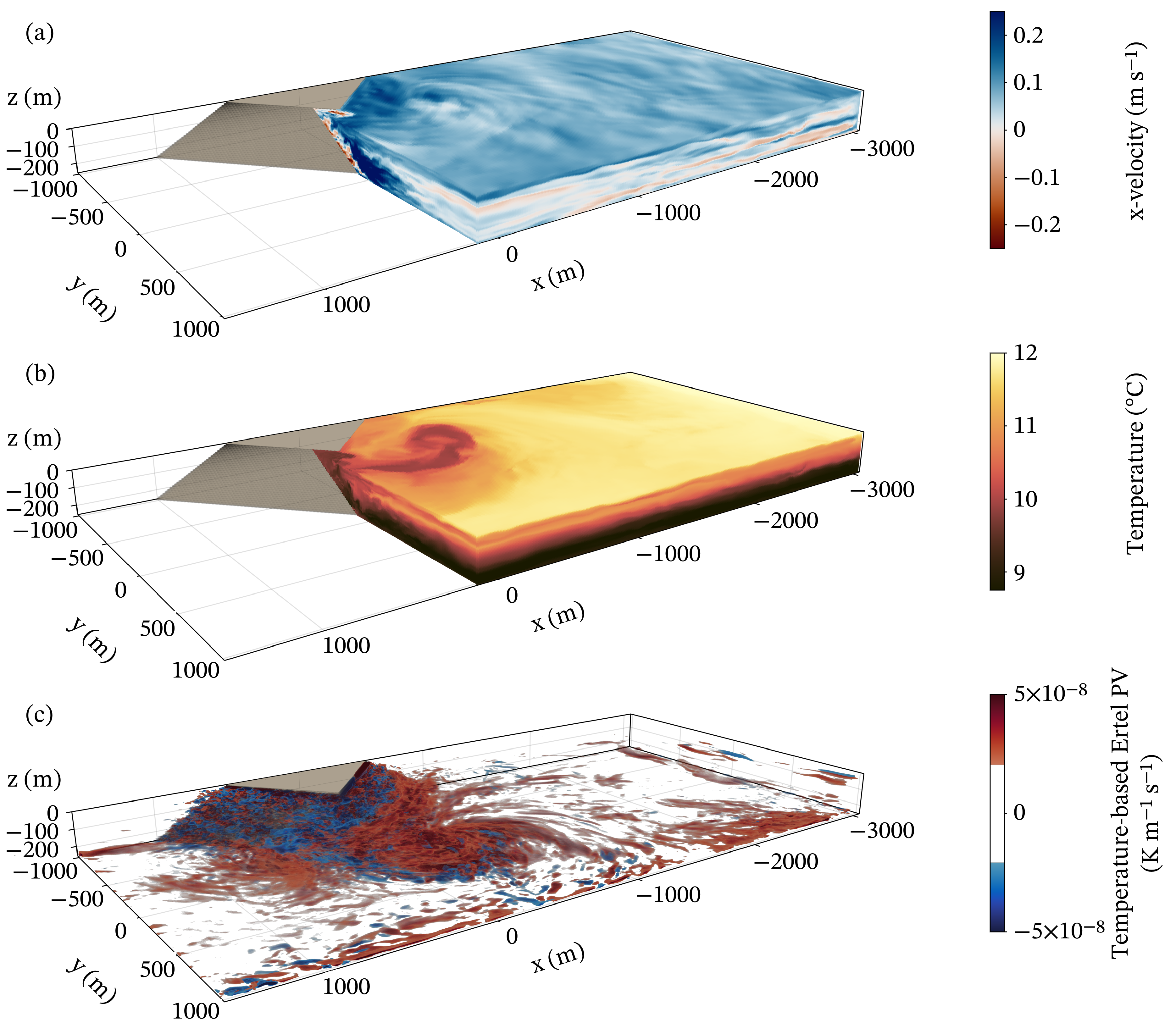}
\caption{(a) Along-channel velocity, (b) temperature, and (c) temperature-based Ertel potential vorticity in a tidally-forced flow past an idealized headland with open boundaries. The tidal flow occurs in the $x$-direction and the snapshot depicts the flow just after the tide has turned to the negative-$x$ direction.}
\label{fig:headland}
\end{figure}

The oscillatory and turbulent flow that results from the set up in listing~\ref{list:headland-les}, simulated on a single Nvidia H100 GPU, is visualized in figure~\ref{fig:headland}.
Figure~\ref{fig:headland}a shows the along-channel velocity field, revealing the formation of a turbulent wake downstream of the headland as the tide reverses.
Figure~\ref{fig:headland}b shows temperature, illustrating how tidal flow over the sloping bathymetry mixes the initially stratified water column.
Figure~\ref{fig:headland}c shows temperature-based Ertel potential vorticity,
\beq
\Pi_T = \left(\bnabla \times \bu + f\bzh\right) \bcdot \bnabla T \com
\label{eq:temperature-pv}
\eeq
where $f$ is the Coriolis parameter and $\bzh$ is the vertical unit vector.
The diagnostic has units $\mathrm{K\,m^{-1}\,s^{-1}}$ and is computed using the companion package Oceanostics \cite{oceanostics}, which interpolates the vorticity and temperature-gradient components to cell corners before taking their dot product.
The visualization highlights filaments generated by flow separation at the headland tip.
This example demonstrates several capabilities working together: immersed boundaries for complex bathymetry, open boundary conditions for tidal forcing, WENO advection for turbulence-resolving simulation, and composition with an external diagnostics package.
The simulation uses a grid-fitted immersed boundary method, in which solid boundaries are represented by masking cells that lie within the topography.
Oceananigans supports both full-cell (``staircase'') and partial-cell representations of bathymetry within this framework \cite{adcroft1997representation, pacanowski1998transient}.

\subsection{Hydrostatic model with a free surface}
\label{sec:hydrostatic-model-equations}

The \code{HydrostaticFreeSurfaceModel} solves the \textit{hydrostatic}, rotating Boussinesq equations with a free surface.
The hydrostatic approximation, inherent to the \code{HydrostaticFreeSurfaceModel}, means that the vertical momentum equation used by \code{NonhydrostaticModel}, $\bzh \bcdot \eqref{nonhydrotatic-momentum-equations}$, is replaced by a statement of hydrostatic balance,
\beq
\d_z p = b \com
\eeq
while the vertical velocity is obtained diagnostically from the continuity equation,
\beq \label{hydrostatic-continuity}
\d_z w = - \nablah \bcdot \bu_h \per
\eeq
(Above we assumed that gravity points down, i.e., $\b{\hat g} = - \bzh$ in~\eqref{nonhydrotatic-momentum-equations}.)
As a result, time-stepping the \code{HydrostaticFreeSurfaceModel} does not require solving a three-dimensional Poisson equation for pressure.
Moreover, the \code{HydrostaticFreeSurfaceModel} introduces a free surface displacement $\eta$, which obeys the linearized equation
\beq \label{free-surface-evolution}
\d_t \eta = w |_{z=0} \per
\eeq
Equation~\eqref{free-surface-evolution} replaces the rigid-lid impenetrability condition $w |_{z=0} = 0$ typically applied at top boundaries in the \code{NonhydrostaticModel}.
The numerical algorithms and computational performance of the \code{HydrostaticFreeSurfaceModel} are described in more detail by \citeA{silvestri2025gpu}.

In the \code{HydrostaticFreeSurfaceModel}, the horizontal velocity $\bu_h = u \bxh + v \byh$ evolves according to
\beq \label{hydrostatic-equations}
\d_t \bu_h =
- \nablah p
- \underbrace{\strut g \nablah \eta}_{\text{free surface}}
- \ \underbrace{\strut \left ( \bu \bcdot \bnabla \right ) \bu_h}_{\substack{\text{momentum}\\\text{advection}}}
\ - \ \underbrace{\strut \b{f} \times \bu}_{\text{Coriolis}}
- \  \underbrace{\strut \bnabla \bcdot \b{\tau}}_{\text{closure}}
\  + \underbrace{\strut \b{F}_{uh}}_{\text{forcing}} \com
\eeq
where $p$ is the hydrostatic kinematic pressure anomaly, $\eta$ is the free surface displacement, $\bu = u \bxh + v \byh + w \bzh$ is the three-dimensional velocity, $\b{f}$ is the background vorticity associated with a rotating frame of reference, $\b{\tau}$ is the stress associated with subgrid turbulent horizontal momentum transport, and $\b{F}_{uh}$ is a body force.
As described in~\ref{sec:time-discretization} and by \citeA{silvestri2026rungekutta}, equation~\eqref{hydrostatic-equations} may be integrated with a predictor-corrector split-explicit method that uses short substeps for the ``barotropic mode'' (the fast component of the free surface displacement $\eta$ and vertically-integrated horizontal momentum) and a much slower step for the three-dimensional part of the solution.
Implicit and fully explicit methods for the combined time-integration of the momentum \eqref{hydrostatic-equations} and the free surface \eqref{free-surface-equation} are also available.

Momentum advection in \code{HydrostaticFreeSurfaceModel} can be formulated in three ways: \emph{(i)}~in the same flux form used by \code{NonhydrostaticModel},
\beq \label{hydrostatic-flux-advection}
\left ( \bu \bcdot \bnabla \right ) \bu_h = \bnabla \bcdot \left ( \bu \, \bu_h \right ) \com
\eeq
\emph{(ii)}~in a standard ``vector invariant'' form that facilitates simulations on curvilinear grids \cite{adcroft2004implementation},
\beq
\left ( \bu \bcdot \bnabla \right ) \bu_h = \zeta \bzh \times \bu_h + w \, \d_z \bu_h + \nablah \left(\half | \bu_h |^2\right)\com
\eeq
and \emph{(iii)}~a modified vector invariant form,
\beq \label{hydrostatic-weno-vi-advection}
\left ( \bu \bcdot \bnabla \right ) \bu_h = \zeta \bzh \times \bu_h - \bu_h \, \d_z w + \d_z \left ( w \, \bu_h \right ) + \nablah \left(\half | \bu_h |^2\right) \com
\eeq
that is used for a WENO-based vector invariant scheme \cite{silvestri2024new}. %
The WENO-based vector invariant scheme is particularly well-suited for simulating geophysical turbulence subject to strong Coriolis forces,
because it selectively dissipates enstrophy and the variance of divergence and does not require an explicit turbulence closure for numerical stability \cite{silvestri2024new}.
An optional, more expensive energy-conserving upwind reconstruction pairs the vorticity fluxes symmetrically so that they do not change the resolved kinetic energy.
(\citeA{zhang2025weno} show how the WENO vector invariant methodology may be translated to apply to potential vorticity in a layered formulation, and thereby dissipate potential vorticity.)
This contrasts the WENO vector invariant scheme with second-order flux form or standard vector invariant schemes, which produce oscillatory errors and must be paired with an explicit turbulence closure, and with a flux form WENO scheme, which targets the dissipation of kinetic energy rather than enstrophy.

Tracer evolution in the \code{HydrostaticFreeSurfaceModel} is governed by the conservation law
\beq
\d_t c \; = \;
- \underbrace{\strut \bnabla \bcdot \left [ \left ( \bu + \bubgc \right ) c \right ] + c \bnabla \bcdot \bu_p }_{\text{tracer advection}}
\quad - \quad \underbrace{\strut \bnabla \bcdot \b{J}_c}_{\text{closure}}
\quad + \quad \underbrace{\strut S_c}_{\text{biogeochemistry}}
\quad + \quad \underbrace{\strut F_c}_{\text{forcing}} \com
\label{eq:hydrostatic-tracer-evolution}
\eeq
which is identical to the \code{NonhydrostaticModel} tracer equation~\eqref{nonhydrostatic-tracer-equation}, except that background fields are not supported.
\code{HydrostaticFreeSurfaceModel} also supports prescribing the velocity field $\bu$ in~\eqref{eq:hydrostatic-tracer-evolution}, enabling cheap tracer advection and Lagrangian particle trajectory simulations.

\begin{figure}[!h]
\centering
\includegraphics[width=1.0\textwidth]{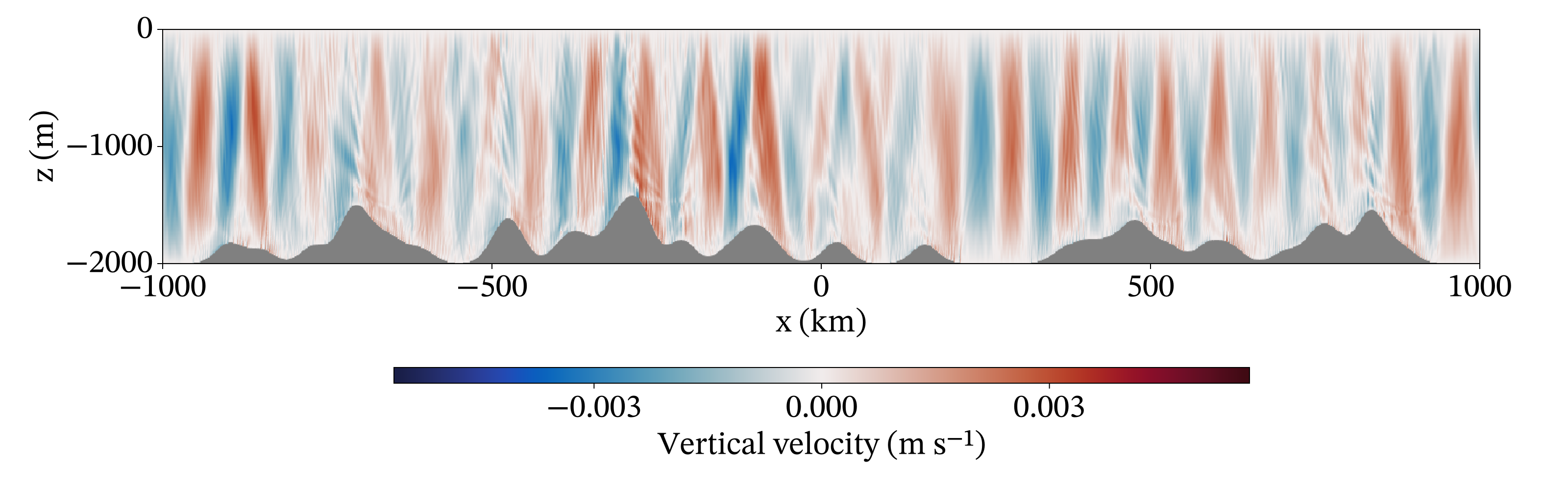}
\caption{Vertical velocity of an internal wave field excited by tidally-forced stratified flow over a superposition of randomly-positioned Gaussian seamounts, after 16 tidal periods.}
\label{fig:internal-tide}
\end{figure}

Listing~\ref{list:internal-tide} implements a two-dimensional simulation of tidally-forced stratified flow over a series of randomly-positioned Gaussian seamounts, using flux form WENO advection schemes to dissipate kinetic energy and tracer variance, as appropriate for a two-dimensional simulation dominated by wave motion.
In this simulation, the horizontal flow associated with tidal forcing is forced upwards over the seamounts, perturbing the stable, vertically-stratified buoyancy distribution and forcing the development of internal inertia-gravity waves, sometimes called ``internal tides'' \cite{garrett2007internal}.
The vertical velocity is visualized in figure~\ref{fig:internal-tide}, showing that the tidal signal is dominated by ``mode-one'', whose structure extends smoothly from the top to the bottom of the domain.

\begin{jllisting}[float, caption={Two-dimensional simulation of tidally-forced stratified flow over a superposition of randomly-positioned Gaussian seamounts. Results are shown in Figure \ref{fig:internal-tide}.}, label={list:internal-tide}]
using Oceananigans, Oceananigans.Units 

grid = RectilinearGrid(size = (2000, 200), halo = (4, 4),
    x = (-1000kilometers, 1000kilometers),
    z = (-2kilometers, 0),
    topology = (Periodic, Flat, Bounded))

h₀ = 100          # typical mountain height (m)
δ  = 20kilometers # mountain width (m)
seamounts = 42
W = grid.Lx - 4δ
x₀ = W .* (rand(seamounts) .- 1/2) # mountains' positions ∈ [-Lx/2+2δ, Lx/2-2δ]
h  = h₀ .* (1 .+ rand(seamounts))  # mountains' heights ∈ [h₀, 2h₀]

bottom(x) = -grid.Lz + sum(h[s] * exp(-(x - x₀[s])^2 / 2δ^2) for s = 1:seamounts)
grid = ImmersedBoundaryGrid(grid, GridFittedBottom(bottom))

T₂ = 12.421hours # period of M₂ tide constituent 
@inline tidal_forcing(x, z, t, p) = p.U₂ * 2π / p.T₂ * sin(2π / p.T₂ * t)
u_forcing = Forcing(tidal_forcing, parameters=(; U₂=0.1, T₂=T₂))

model = HydrostaticFreeSurfaceModel(; grid, tracers=:b, buoyancy=BuoyancyTracer(),
    momentum_advection=WENO(), tracer_advection=WENO(),
    forcing=(; u=u_forcing))

bᵢ(x, z) = 1e-5 * z
set!(model, b=bᵢ)
\end{jllisting}

\subsubsection{Vertical mixing and mesoscale turbulence parameterizations}
\label{sec:mixing-parameterizations}

Similar to the \code{NonhydrostaticModel}, the closure tracer fluxes $\b{J}_c$ may be three-dimensional\-ly, horizontally, or vertically Laplacian or biharmonic.
In addition, a closure is provided that computes both isopycnal ``symmetric'' diffusive tracer fluxes \cite{redi1982oceanic} and skew diffusive tracer fluxes \cite{gent1990isopycnal, griffies1998gent}, which are typically used to parameterize stirring and vertical form stress, respectively, associated with unresolved mesoscale turbulence \cite<e.g.>{young2012exact}.
The symmetric contribution (often called ``Redi'') and skew contribution (often called ``GM'' after Gent--McWilliams) to the tracer flux may be tapered in the presence of a large isopycnal slope.
The GM flux may be formulated as tracer advection, permitting combination with the other velocity components $\bu$ and $\bubgc$, but with the downside of invoking additional gradients of the isopycnal slope \cite{griffies1998gent}.

Oceananigans provides a number of closures that parameterize vertical fluxes of tracers and momentum associated with unresolved turbulence with scales from O(1)--O(100) meters.
Depending on the parameterization, the evolution of auxiliary tracers like turbulent kinetic energy and the turbulent kinetic energy dissipation rate may also be simulated.
Vertical mixing parameterizations are useful for hydrostatic simulations where vertical mixing is otherwise unresolved due to a coarse horizontal grid spacing \cite<see>[and references therein]{large1994oceanic, vanroekel2018kpp}.
For example, in such regional and global configurations, horizontal grid spacing typically varies from $O(100 \, \mathrm{m})$ to $O(100 \, \mathrm{km})$.

In addition to providing a number of parameterizations useful for hydrostatic ocean modeling on coarse grids, Oceananigans also provides an interface for implementing new parameterizations.
This interface, which is invoked by extending the default behavior of closure functions with custom implementations using multiple dispatch, is illustrated in listing~\ref{list:pp-closure} with the implementation of the Pacanowski--Philander Richardson-number-based vertical diffusivity with a convective adjustment modification.
A complete reference to the closure interface is given in \href{https://clima.github.io/OceananigansDocumentation/stable/developer_docs/turbulence_closures/}{the developer documentation}.

\begin{jllisting}[float, caption={A partial demonstration of the implementation of the Pacanowski--Philander Richardson-number-based vertical diffusivity as a turbulence closure in Oceananigans. The implementation extends \texttt{compute\_closure\_fields!}, leveraging the new closure type, and defines a ``kernel'' that runs on both CPU and GPU using KernelAbstractions syntax. The implementation illustrates the pattern of using the new type to both extend methods as well as to hold parameters of the new component. The full reference implementation, including time-discretization options and show methods, is given in the Oceananigans developer documentation. The closure extends the original Pacanowski--Philander model with a convective-adjustment branch with $(\nu_{\mathrm{conv}},\kappa_{\mathrm{conv}})$ when $N^2 < 0$, mirroring the pattern used by the built-in \code{RiBasedVerticalDiffusivity}.}, label={list:pp-closure}]
using Oceananigans.TurbulenceClosures
using Oceananigans.TurbulenceClosures: AbstractScalarDiffusivity, VerticalFormulation
using Oceananigans.Operators: ℑxᶜᵃᵃ, ℑyᵃᶜᵃ, ∂zᶠᶜᶠ, ∂zᶜᶠᶠ
using Oceananigans.BuoyancyFormulations: ∂z_b
using KernelAbstractions: @kernel, @index

struct PacanowskiPhilanderDiffusivity{TD, FT} <: AbstractScalarDiffusivity{TD, VerticalFormulation, 1}
    ν₀ :: FT;  ν₁ :: FT;  κ₀ :: FT
    ν_conv :: FT;  κ_conv :: FT
    c :: FT;  n :: FT
end

function compute_closure_fields!(diffs, closure::PacanowskiPhilanderDiffusivity, model)
    launch!(architecture(model), model.grid, :xyz, _pp_diffusivities!,
            diffs, model.grid, closure, model.velocities, model.tracers, model.buoyancy)
    return nothing        
end

# Vertical shear squared at (Center, Center, Face), built from the same
# staggered-grid operators used to compute velocity gradients.
@inline ϕ²(i, j, k, grid, ϕ, args...) = ϕ(i, j, k, grid, args...)^2
@inline shear_squaredᶜᶜᶠ(i, j, k, grid, U)=ℑxᶜᵃᵃ(i, j, k, grid, ϕ², ∂zᶠᶜᶠ, U.u) +
    ℑyᵃᶜᵃ(i, j, k, grid, ϕ², ∂zᶜᶠᶠ, U.v)

@kernel function _pp_diffusivities!(diffs, grid, closure, velocities, tracers, buoyancy)
    i, j, k = @index(Global, NTuple)
    N²  = ∂z_b(i, j, k, grid, buoyancy, tracers)
    S²=max(shear_squaredᶜᶜᶠ(i, j, k, grid, velocities), convert(eltype(grid), 1e-10))
    Ri  = max(zero(grid), N²) / S²
    den = 1 + closure.c * Ri
    νz_s = closure.ν₀ + closure.ν₁ / den^closure.n
    κz_s = closure.κ₀ + closure.ν₀ / den + closure.ν₁ / den^(closure.n + 1)
    conv = N² < 0
    diffs.νz[i, j, k] = ifelse(conv, closure.ν_conv, νz_s)
    diffs.κz[i, j, k] = ifelse(conv, closure.κ_conv, κz_s)
end
\end{jllisting}

\begin{figure}[htp]
\centering
\includegraphics[width=1.0\textwidth]{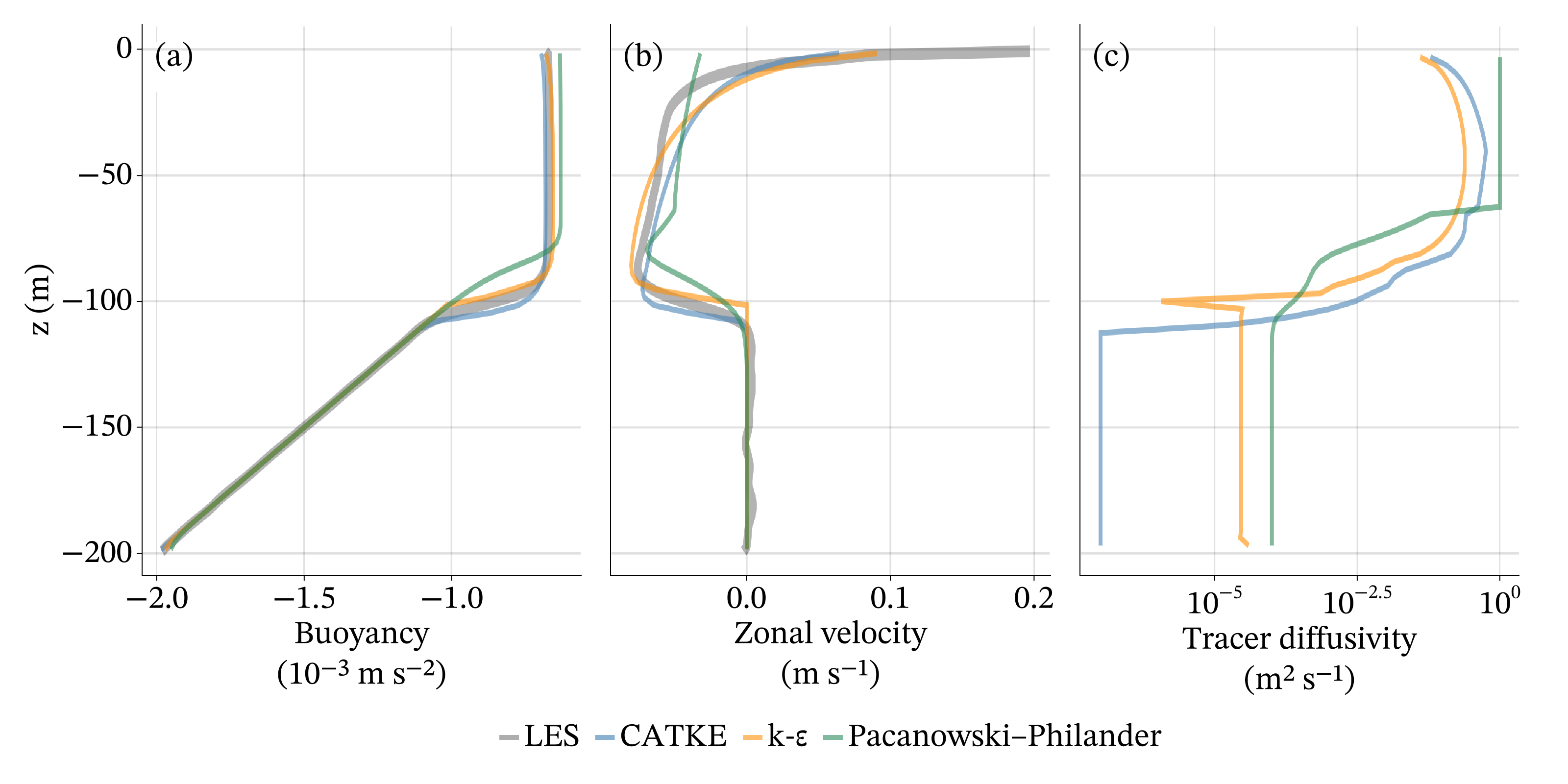}
\caption{Results from four Oceananigans simulations. Three of the simulations are column models using the CATKE, $k$-$\epsilon$, and Pacanowski--Philander vertical mixing parameterizations, compared against a 3D nonhydrostatic large-eddy simulation (LES) reference of the same wind/buoyancy-forced boundary layer with surface waves ($\mathrm{La}=0.3$). The LES (black) is run on a $128^3$ grid with WENO(9) advection; the parameterizations (colored) are evaluated on a 32-level vertical column with $\Delta t = 5$~min. Panels: \emph{(a)} buoyancy $b(z)$, \emph{(b)} zonal velocity $u(z)$, \emph{(c)} tracer eddy diffusivity $\kappa(z)$ (log scale). The parameterizations are implemented as described in Listings~\ref{list:parameterizations} and~\ref{list:pp-closure}.}
\label{fig:parameterizations}
\end{figure}

To illustrate vertical mixing closures in action, listing~\ref{list:parameterizations} implements a simulation of convectively- and wind-driven vertical mixing using both a reference large eddy simulation and in a single column model context.
We compare the LES with three parameterizations: CATKE \cite{wagner2023catke}, which has one additional equation for the evolution of turbulent kinetic energy (TKE); $k$-$\epsilon$ \cite{umlauf2005second}, which has two additional equations for both TKE and TKE dissipation; and our custom implementation of \citeA{pacanowski1981parameterization}.
Figure~\ref{fig:parameterizations} plots the result, showing how $k$-$\epsilon$ mixes less than CATKE while the Pacanowski--Philander scheme produces smoother diffusivity profiles set by the local shear and stratification.
The discrepancy in mixing rates between CATKE and $k$-$\epsilon$ is likely due to differences in how the models are calibrated.
While all of CATKE's parameters are jointly calibrated to 35 large eddy simulations (LES) that include surface wave effects \cite{wagner2023catke}, $k$-$\epsilon$ parameters are calibrated one-by-one by referencing laboratory experiments and observations of increasing complexity \cite{umlauf2003generic}.
Both CATKE and $k$-$\epsilon$ may be optionally substepped, a useful performance optimization for simulations on coarse grids that can otherwise accommodate relatively long baroclinic time steps.

\begin{jllisting}[float, caption={Code for comparing three vertical mixing parameterizations (CATKE, $k$-$\epsilon$, and Pacanowski--Philander) against a 3D nonhydrostatic LES ``reference'' in the evolution of an initially linearly stratified boundary layer subjected to stationary surface fluxes of buoyancy and momentum, plus a wave-aligned Stokes drift with Langmuir number $\mathrm{La}=0.3$. Results are shown in Figure~\ref{fig:parameterizations}. The argument \texttt{pp} is a fully constructed Pacanowski--Philander closure with the field allocation and accessor methods described in the developer documentation; listing~\ref{list:pp-closure} shows its diffusivity kernel. The LES uses \texttt{boundary\_layer\_simulation(nothing, GPU(); les=true)}.}, label={list:parameterizations}]
using Oceananigans
using Oceananigans.Units
using Oceananigans.StokesDrifts: UniformStokesDrift
using Oceananigans.TurbulenceClosures: CATKEVerticalDiffusivity, TKEDissipationVerticalDiffusivity

function boundary_layer_simulation(closure, arch=CPU(); les=false, La=0.3, λ=60,
    N²=1e-5, Jb=1e-7, τx=-5e-4, f=1e-4)

    u_bcs = FieldBoundaryConditions(top=FluxBoundaryCondition(τx))
    b_bcs = FieldBoundaryConditions(top=FluxBoundaryCondition(Jb))
    model_kw = (; closure, coriolis = FPlane(; f),
                buoyancy = BuoyancyTracer(), tracers=:b,
                boundary_conditions = (u=u_bcs, b=b_bcs))

    if les
        x = y = (-200, 200)
        grid = RectilinearGrid(arch, size=(128, 128, 128); x, y, z=(-200, 0))
        advection = WENO(order=9)

        # Aligned wind/wave Stokes drift: u_star = sqrt(|τx|), uˢ_surf = u_star / La²,
        # decay scale 1/(2k) where k = 2π/λ is the wavenumber.
        @inline ∂z_uˢ(z, t, p) = 2 * p.k * p.uˢ * exp(2 * p.k * z)
        stokes_drift = UniformStokesDrift(grid; ∂z_uˢ,
            parameters = (k = 2π/λ, uˢ = sqrt(abs(τx)) / La^2))

        model = NonhydrostaticModel(grid; advection, stokes_drift, model_kw...)
        set!(model, b = (x, y, z) -> N² * z + 1e-6 * rand())
    else
        grid=RectilinearGrid(arch, size=32, z=(-200, 0), topology=(Flat, Flat, Bounded))
        model = HydrostaticFreeSurfaceModel(grid; model_kw...)
        set!(model, b = z -> N² * z)
    end

    simulation = Simulation(model; Δt=5minute, stop_time=2days)
    les && conjure_time_step_wizard!(simulation, cfl=0.7)

    run!(simulation)
    return simulation
end

# pp is an instance of the complete custom closure illustrated in listing 9.
function compare_vertical_mixing(pp)
    catke = boundary_layer_simulation(CATKEVerticalDiffusivity())
    kϵ = boundary_layer_simulation(TKEDissipationVerticalDiffusivity())
    pacanowski_philander = boundary_layer_simulation(pp)
    return (; catke, kϵ, pacanowski_philander)
end
\end{jllisting}

\subsubsection{Ocean simulations on global spherical shell grids}

\begin{figure}[htp]
\centering
\includegraphics[width=1.0\textwidth]{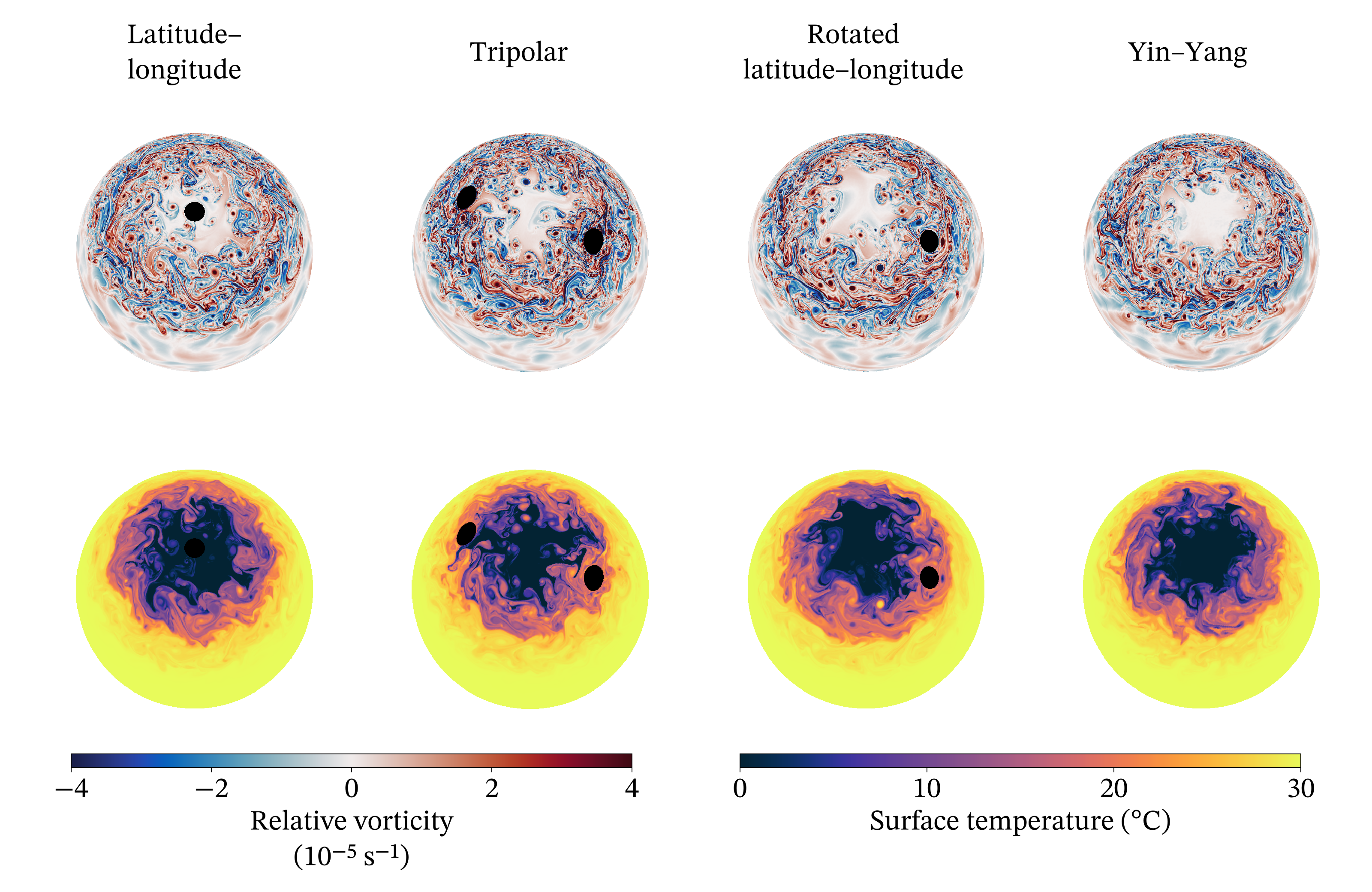}
\caption{Relative vorticity (top row) and surface temperature (bottom row) after 180 days of baroclinic instability on four grids. From left to right: latitude--longitude, tripolar with cylindrical islands centered on the two northern grid singularities, rotated latitude--longitude with a grid pole at $55^\circ$N, $70^\circ$E, and Yin--Yang. Black regions denote islands or excluded polar caps. Listing~\ref{list:near-global} illustrates the setup for all four grids.}
\label{fig:near-global}
\end{figure}

The \code{HydrostaticFreeSurfaceModel} can be used to simulate regional or global ocean circulation on rectilinear grids, latitude-longitude grids, and the tripolar grid~\cite{murray1996explicit} to cover the entirety of Earth's global ocean with its current continental configuration.
The tripolar grid avoids the coordinate singularity at the geographic North Pole by replacing it with two singularities located over land, typically in northern Canada and Siberia. 
This is achieved by mapping a family of concentric ellipses obtained from a stereographic projection onto the sphere. The singularities correspond to the foci of the ellipses and can therefore be masked within land regions.
As a consequence, the grid contains a northern ``fold'' or seam connecting the two artificial poles. 
Oceananigans supports two fold configurations: the seam can lie either at \code{(Face, Face)} locations (tracer cell corners) or at \code{(Face, Center)} locations (tracer cell edges).
Across this fold, grid points on one side are mapped to corresponding indices on the opposite side of the seam using a zipper boundary condition following \citeA{murray1996explicit}, which pairs fields across the seam while accounting for the appropriate sign changes in vector components.
A cubed sphere grid \cite{sadourny1972conservative, ronchi1996cubed, ranvcic1996global, adcroft2004implementation} that covers the entire sphere and is therefore useful for aquaplanet simulations, is also implemented and currently being validated.
To illustrate simulations on global and near-global grids, we simulate baroclinic instability on four spherical grids: a latitude-longitude grid, a tripolar grid with islands placed over the north pole singularities, a ``rotated'' latitude-longitude grid wherein the polar grid singularities are rotated from the geographic north pole to $55^\circ$ N, $70^\circ$ E, corresponding to one of the tripolar grid's two north poles, and a Yin--Yang grid.
Each simulation runs on a single Nvidia H100 GPU.
A portion of the code that produces the simulation is given in listing~\ref{list:near-global} and the results are visualized in figure~\ref{fig:near-global}.

\begin{jllisting}[float, caption={Global and near-global simulations of baroclinic instability on four grids.}, label={list:near-global}]
arch = GPU()
Nx, Ny, Nz = size = (4 * 360, 4 * 170, 10)
halo = (7, 7, 7)
H = 3000
latitude, longitude, z = (-85, 85), (0, 360), (-H, 0)

lat_lon_grid = LatitudeLongitudeGrid(arch; size, halo, latitude, longitude, z)
rotated_grid = RotatedLatitudeLongitudeGrid(arch; size, halo, latitude,
    longitude, z, north_pole=(70, 55))
yin_yang_grid = YinYangGrid(arch; size=(4 * 270, 4 * 90, Nz), halo, z)
 
# TripolarGrid with "Gaussian islands" over the two north poles
underlying_tripolar_grid = TripolarGrid(arch; size, halo, z)

dφ, dλ, λ₀, φ₀ = 4, 8, 70, 55
isle(λ, φ) = ((λ - λ₀)^2 / 2dλ^2 + (φ - φ₀)^2 / 2dφ^2) < 1
cylindrical_isles(λ, φ) = H * (isle(λ, φ) + isle(λ - 180, φ) - 1)
tripolar_grid=ImmersedBoundaryGrid(underlying_tripolar_grid, GridFittedBottom(cylindrical_isles))

momentum_advection = WENOVectorInvariant(order=9)
tracer_advection = WENO(order=7)
coriolis = HydrostaticSphericalCoriolis()
buoyancy = SeawaterBuoyancy(equation_of_state=TEOS10EquationOfState())

grid = lat_lon_grid # rotated_grid, tripolar_grid, yin_yang_grid
free_surface = SplitExplicitFreeSurface(grid, substeps=60,
    extend_halos=!(grid isa YinYangGrid))
model = HydrostaticFreeSurfaceModel(; grid, momentum_advection, tracer_advection,
    coriolis, free_surface, buoyancy, tracers=(:T, :S))

Tᵢ(λ, φ, z) = 30 * (1 - tanh((abs(φ) - 45) / 8)) / 2 + rand()
Sᵢ(λ, φ, z) = 28 - 5e-3 * z + rand()
set!(model, T=Tᵢ, S=Sᵢ)

simulation = Simulation(model, Δt=2minutes, stop_time=180days)
grid isa YinYangGrid && add_conservation_fixer!(simulation)
\end{jllisting}

\subsubsection{Realistic ocean--sea ice simulations beneath prescribed atmospheric states with NumericalEarth}
\label{sec:clima-ocean}

The coupled modeling package NumericalEarth \cite{NumericalEarth-zenodo} implements an Oceananigans-based framework for coupled modeling including model integration, diagnostics, flux, interfacial state computations --- functionality that is not provided by Oceananigans itself.
In NumericalEarth, turbulent interfacial fluxes are computed using Monin--Obukhov similarity theory \cite{monin1954basic} following \citeA{edson2014exchange} for air-sea fluxes and \citeA{grachev2007sheba} for air-ice fluxes.
NumericalEarth additionally provides utilities for downloading and interfacing with JRA55 reanalysis data \cite{tsujino2018jra}, building grids based on Earth bathymetry and initializing simulations from datasets including the Estimating the Circulation and Climate of the Ocean (ECCO) state estimate \cite{forget2015ecco}, Mercator Ocean's Global Ocean Reanalysis (GLORYS), and EN4 \cite{good2013en4}.

\begin{figure}[!h]
\centering
\includegraphics[width=1.0\textwidth]{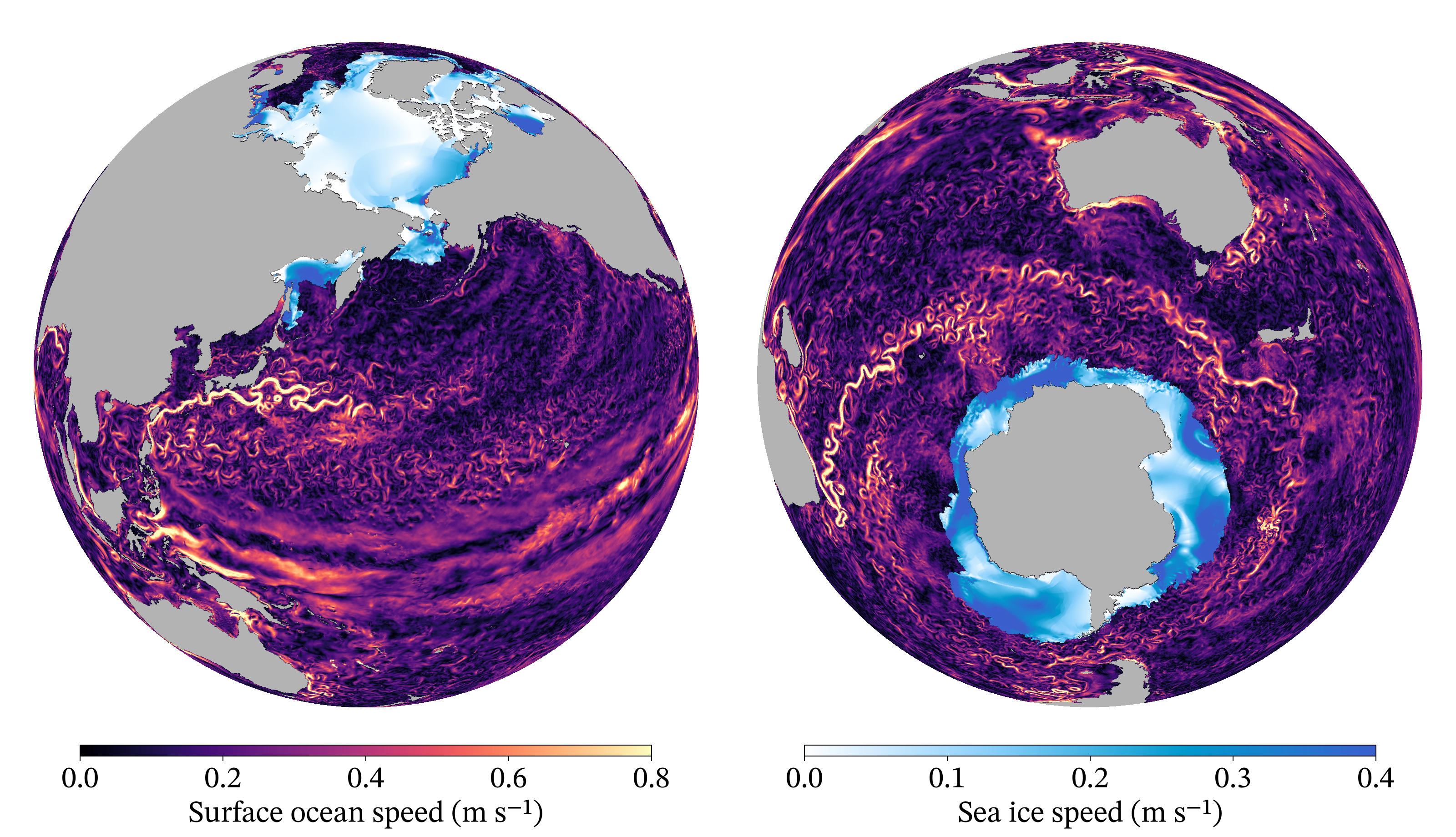}
\caption{Ocean surface and sea ice speed in a coupled ocean--sea ice simulation without tides on a 1/12th degree tripolar grid \protect\cite{murray1996explicit}, viewed from two perspectives. The snapshot is taken on 1 May 2015, after 120 simulated days. Sea ice speed is shown where ice concentration is at least $10^{-3}$.}
\label{fig:clima-ocean}
\end{figure}

We illustrate NumericalEarth's capabilities with a coupled ocean and sea ice simulation on a tripolar grid with a nominal horizontal grid spacing of 1/12$^\circ$ and $4320 \times 2040 \times 100$ cells, extending from 80$^\circ$S to the northern grid poles.
The ocean component uses Oceananigans' \code{HydrostaticFreeSurfaceModel}, \code{WENOVectorInvariant} momentum advection, seventh-order WENO tracer advection, CATKE vertical mixing, the \code{TEOS10EquationOfState}, and ETOPO2022 bathymetry.
The sea ice component uses ClimaSeaIce, with a zero-layer thermodynamic formulation, constant heat conductivity, incremental-remapping advection, and a viscoplastic rheology integrated with 120 substeps using the modified elastic-viscous-plastic solver of \citeA{kimmritz2016adaptive}.
The simulation uses prescribed atmospheric states and radiation from ERA5 \cite{hersbach2020era5} and river runoff from JRA55-do \cite{tsujino2018jra}.
Freshwater exchange is represented by volume fluxes through the nonlinear free surface.
Ocean temperature and salinity and sea ice thickness and concentration are initialized from GLORYS12 \cite{lellouche2021glorys} on 1 January 2015, with the ocean initially at rest.
Listing~\ref{list:clima-ocean} illustrates the coupled configuration.
Figure~\ref{fig:clima-ocean} shows ocean surface speed (black to yellow) and sea ice speed (blue) on 1 May 2015, after 120 simulated days without tidal forcing.
The simulation uses four Nvidia GH200 GPUs, with an adaptive time step targeting a horizontal CFL number of 0.5 and capped at 10~min.
The 120-day integration took 12.7 hours including startup, equivalent to approximately 0.62 simulated years per wall-clock day (SYPD).
Excluding startup, throughput was 0.68 SYPD, including surface output every six simulated hours, daily sea surface height output, atmospheric data reloads, and checkpoints every five simulated days.
For more information about Oceananigans performance on multiple GPUs, see \citeA{silvestri2025gpu}.

\begin{jllisting}[float, caption={Selected configuration code for the 1/12$^\circ$ coupled simulation in figure~\ref{fig:clima-ocean}, using ERA5 atmospheric forcing, JRA55-do runoff, and GLORYS12 initial conditions. This excerpt uses the figure run's Oceananigans 0.113.6 and NumericalEarth 0.8.2 development configuration. Imports, bathymetry editing, missing-data handling, time-stepping callbacks, and output are omitted.}, label={list:clima-ocean}]
arch = NCCLDistributed(GPU(); partition = Partition(1, 4)) # 4 GPUs
Nx, Ny, Nz = 4320, 2040, 100 # 1/12th degree
z = ExponentialDiscretization(Nz, -6000, 0; scale=1100, mutable=true) # z*
underlying_grid = TripolarGrid(arch; size=(Nx, Ny, Nz), halo=(7, 7, 4), z)
bottom_height = regrid_bathymetry(underlying_grid) # ETOPO2022
# Production bathymetry editing omitted here.
grid = ImmersedBoundaryGrid(underlying_grid, GridFittedBottom(bottom_height))

start_date, end_date = DateTime(2015, 1, 1), DateTime(2015, 5, 2)
max_Δt = 10minutes
land = JRA55PrescribedLand(grid; dataset=MultiYearJRA55(), start_date, end_date) # runoff

free_surface = NumericalEarth.Oceans.default_free_surface(grid; fixed_Δt=max_Δt)
advection_time_discretization = AdaptiveVerticallyImplicitDiscretization(cfl=0.5)
tracer_advection=WENO(order=7; time_discretization=advection_time_discretization)
                        
ocean = ocean_simulation(grid; free_surface, tracer_advection,
    river_routing=land.river_routing)

dynamics = NumericalEarth.SeaIces.sea_ice_dynamics(grid, ocean;
    rheology = ElastoViscoPlasticRheology(max_relaxation_parameter=120),
    solver = SplitExplicitSolver(grid; substeps=120))
sea_ice = sea_ice_simulation(grid, ocean; dynamics,
    advection=IncrementalRemapping(), timestepper=:ForwardEuler)

glorys=MetadataSet(:temperature, :salinity, :sea_ice_thickness, :sea_ice_concentration;
    dataset=GLORYSDaily(), date=start_date)
set!(ocean.model, glorys)   # the ocean starts at rest
set!(sea_ice.model, glorys)
# Production initialization replaces missing ice values with zero.

atmosphere = ERA5PrescribedAtmosphere(arch; start_date, end_date,
    pressure=:mean_sea_level_pressure)
ocean_surface = SurfaceRadiationProperties(albedo=LatitudeDependentAlbedo())
radiation = ERA5PrescribedRadiation(arch; start_date, end_date, ocean_surface)
coupled_model = OceanSeaIceModel(ocean, sea_ice;
    atmosphere, land, radiation, frazil_formation_depth=50)
\end{jllisting}

\subsection{Synthesis of modeling choices for different problems}

Contemplating the computation of ocean-flavored fluid dynamics, the first choice users face is between the nonhydrostatic and hydrostatic models.
The \code{NonhydrostaticModel} is suited to small-scale dynamics --- direct numerical simulations (DNS) and large eddy simulations (LES) of the surface boundary layer, bottom boundary layer, and similar problems --- where grid spacing is typically finer than 10--30~m \cite{marshall1997finite}.
Due to the availability of fast pressure solvers, rectilinear grids (either directly or immersed in bathymetry) are preferred with \code{NonhydrostaticModel}, though latitude-longitude grids are also supported if needed.
The \code{HydrostaticFreeSurfaceModel} targets larger-scale regional to global problems that usually require curvilinear grids (latitude-longitude or tripolar), though rectilinear grids are also supported and can offer performance advantages, especially in distributed configurations.
Additional physics such as Coriolis forces, buoyancy forces, Stokes drift terms, boundary conditions, and forcing terms are then selected according to the computational design.

Numerical and parameterization choices depend on the problem.
Direct numerical simulations (DNS) require \code{ScalarDiffusivity} to model molecular viscosity and diffusion, paired with centered advection. The cylinder DNS examples use \code{Centered(order=4)} and \code{ScalarDiffusivity} with \code{stencil\_order=4}; the cabbeling example uses second-order centered advection and diffusion.
For examples of direct numerical simulation, see listings~\ref{list:flow-around-cylinder} and \ref{list:cabbeling}.
For large eddy simulation, implicit LES utilizing 9th-order WENO advection scheme and no explicit closure produces simulations that minimize numerical dissipation and permits the coarsest grids (listings~\ref{list:eady-les} and \ref{list:headland-les}).
Lower-order WENO schemes (7th or 5th) tend to dissipate more than necessary for well-resolved dynamics, while solutions generated with higher-order WENO schemes (11th and higher) may exhibit spurious oscillations due to inadequate dissipation.
Bounds-preserving WENO schemes may be employed to reduce spurious extrema or out-of-bounds values for tracers, such as strictly positive biogeochemical tracers.
When explicit viscosity or diffusivity diagnostics are needed, \code{DynamicSmagorinsky} can be paired with WENO or a centered scheme, though users should test the combination for their application.
For resolving the logarithmic layer near the ocean bottom, even \code{DynamicSmagorinsky} may be insufficient, and \code{SmagorinskyLilly} can be useful.

For idealized hydrostatic problems such as unforced initial-value problems, \code{WENOVectorInvariant} advection is often sufficient on its own.
This is used for the baroclinic instability simulations in listing~\ref{list:near-global}.
When surface forcing is included and vertical turbulence must be parameterized, we recommend \code{CATKEVerticalDiffusivity} with \code{WENOVectorInvariant}.
This is the numerical configuration used for the 1/12th degree simulation in listing~\ref{list:clima-ocean}.
For coarser hydrostatic simulations with unresolved mesoscale eddies, WENO schemes can still be used, albeit at lower order (7th or 5th), and long-running simulations (decades and more) typically additionally require the \code{IsopycnalSkewSymmetricDiffusivity} (also called the Gent-McWilliams and Redi diffusivities) and may benefit from area-scaled biharmonic viscosity.

\section{Discussion}
\label{sec:conclusions}

This paper describes the GPU-based ocean modeling software called ``Oceananigans'' written in the high-level Julia programming language.
Oceananigans provides a productive script-based user interface and introduces a computational design facilitated by powerful operator abstractions which, coupled with the capabilities of GPUs, aims to reduce the computational cost of simulating oceanic motion from millimeter to planetary scales.
Oceananigans thus realizes a particular design for fast, rapidly developed dynamical cores: simple C-grid WENO numerics for turbulence resolving simulations in a high-level language accelerated by GPUs.

Because Oceananigans can run the large eddy simulations presented here on a single GPU and the global ocean--sea-ice simulation on four GPUs, it increases access to these ocean modeling applications.
On larger GPU systems, Oceananigans also enables large ensembles of global ocean simulations.
For example, using 4,800 GPUs on the Perlmutter supercomputer \cite{nersc_perlmutter_architecture}, it is currently possible to complete a 100-member ensemble of century-long global ocean simulations with a horizontal grid spacing of 10 kilometers in 10 days of wall time --- thereby resolving mesoscale turbulent mixing, a prominent bias in ocean models and a fundamental process missing from most climate simulations today.
These new capabilities address uncertainty in ocean heat and carbon uptake in climate projections.

Oceananigans' programmable interface has fostered a growing ecosystem of companion packages, many developed independently outside the core team.
OceanBioME \cite{strong2023oceanbiome} provides biogeochemistry models for ecosystem dynamics and carbon cycle applications, compatible with both hydrostatic and nonhydrostatic simulations.
Oceanostics \cite{oceanostics} implements complex diagnostics in Oceananigans syntax for online and offline analysis of simulations.
OceananigansLagrangianFilter \cite{baker2026lagrangianfilter} performs online and offline Lagrangian filtering of simulation output to separate slow-evolving flow from high-frequency wave motions.
Beyond ocean modeling, Breeze \cite{Breeze-zenodo} provides a limited-area atmosphere model for LES-to-mesoscale simulations, and Terrarium implements differentiable, GPU-accelerated land and ecosystem models, both built on Oceananigans' grids, solvers, and advection schemes.
NumericalEarth supports coupled atmosphere--ocean simulations with Oceananigans and SpeedyWeather \cite{klower2024speedyweather}.
That these packages were developed by independent researchers --- and at early stages of Oceananigans development --- is evidence that Julia's accessibility lowers the barrier to extending and building on ocean modeling software.

Initial work is also under way to couple Oceananigans-based ocean models with prognostic atmosphere models, including the Climate Modeling Alliance atmosphere dynamical core \cite{yatunin2026climate}, SpeedyWeather \cite{klower2024speedyweather}, and the Oceananigans-based atmosphere model Breeze \cite{Breeze-zenodo}.
The DJ4Earth project has used Enzyme \cite{moses2021reverse} and Reactant to develop a differentiable adjoint for Oceananigans \cite{moses2026adjoint}, enabling auto-differentiation to compute the gradients of cost functions that invoke Oceananigans simulations for gradient-based parameter estimation and data assimilation.

Driven by the recent explosion in computationally-intensive applications of machine learning (ML), computational science and technology is advancing at an accelerating rate.
To keep pace --- to continue to use the world's fastest computers, to maintain scientific productivity commensurate with other fields, and to enable the next generation of theory and ML-based parameterizations --- the development of ocean modeling software must also accelerate.
Oceananigans represents a step towards the faster development of ocean modeling software.
But substantial and sustained growth in the developer community is still required for Oceananigans to reach its potential.

\appendix
\counterwithin{equation}{section}
\counterwithin{figure}{section}
\counterwithin{table}{section}
\renewcommand{\theequation}{\Alph{section}\arabic{equation}}
\renewcommand{\thefigure}{\Alph{section}\arabic{figure}}
\renewcommand{\thetable}{\Alph{section}\arabic{table}}

\section{Time stepping and time discretization}
\label{sec:time-discretization}

In this section we describe time stepping methods and time discretization options for the \code{NonhydrostaticModel} and the \code{HydrostaticFreeSurfaceModel}.

\subsection{Time discretization for tracers}

Tracers are stepped forward with similar schemes in the \code{NonhydrostaticModel} and the \code{HydrostaticFreeSurfaceModel}, each of which includes optional implicit treatment of vertical diffusion terms.
Equation~\eqref{nonhydrostatic-tracer-equation} is abstracted into two components,
\beq \label{tracer-with-tendency}
\d_t c = G_c + \d_z \left ( \kappa_z \d_z c \right ) \com
\eeq
where, if specified, $\kappa_z$ is the vertical diffusivity of $c$ to be treated with a \code{VerticallyImplicitTimeDiscretization}, and $G_c$ is the remaining component of the tracer tendency from equation~\ref{nonhydrostatic-tracer-equation}.
(Vertical diffusion treated with an \code{ExplicitTimeDiscretization} is also absorbed into $G_c$.)
We apply a semi-implicit time discretization of vertical diffusion to approximate the integral of~\eqref{tracer-with-tendency} from $t^m$ to $t^{m+1}$,
\beq \label{tracer-time-discretization}
\left ( 1 - \Delta t \, \d_z \, \kappa_z^m \d_z \right ) c^{m+1} = c^m + \int_{t^m}^{t^{m+1}} G_c \di t \com
\eeq
where $\Delta t \defn t^{m+1} - t^m$.
The tendency integral $\int_{t^m}^{t^{m+1}} G_c \di t$ is evaluated using a low-storage third-order Runge-Kutta (RK3) scheme \cite{Williamson1980LowStorageRK, silvestri2026rungekutta}.
For RK3, the indices $m = (1, 2, 3)$ correspond to \textit{substages}, and the integral in~\eqref{tracer-time-discretization} takes the form
\beq \label{rk3}
\frac{1}{\Delta t} \int_{t^m}^{t^{m+1}} G_c \di t \approx \gamma^m G_c^m - \zeta^m G_c^{m-1} \com
\eeq
where $\gamma = (8/15, 5/12, 3/4)$ and $\zeta = (0, 17/60, 5/12)$ for $m = (1, 2, 3)$ respectively.
RK3 requires three evaluations of the tendency $G_c$ per time-step.
RK3 is self-starting because $\zeta^1 = 0$.
Equation~\eqref{tracer-time-discretization} is solved with a tridiagonal algorithm following a second-order spatial discretization of $\d_z \kappa^m_z \d_z c^{m+1}$ at each of RK3's three stages.

\begin{figure}[htp]
\centering
\includegraphics[width=0.9\textwidth]{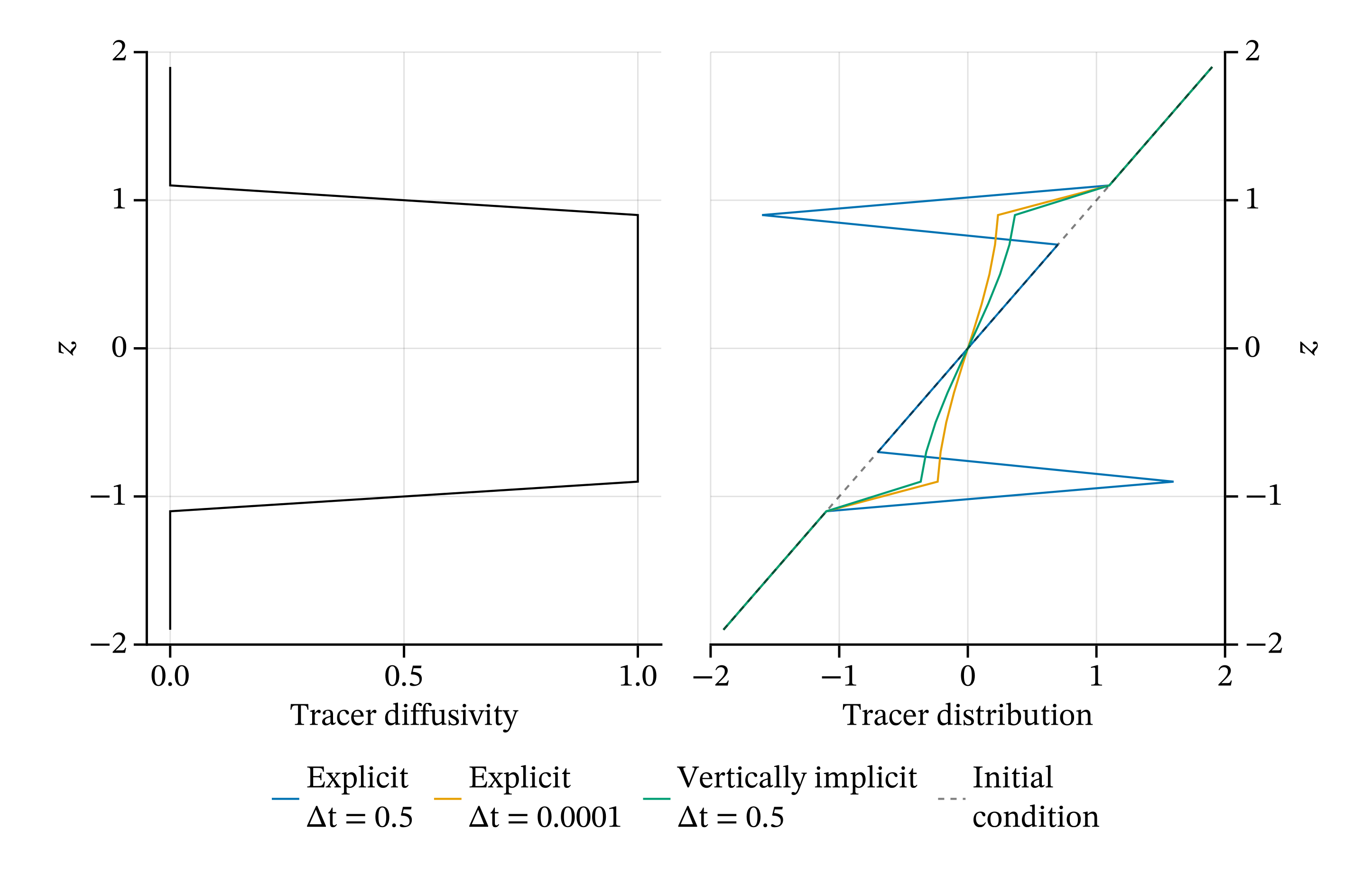}
\caption{\pretolerance=100 Simulations of tracer diffusion by a top hat diffusivity profile using various choices of time-discretization and time-step size. With a long time-step of $\Delta t = 0.5$, \code{ExplicitTimeDiscretization} is unstable while \code{VerticallyImplicitTimeDiscretization} is stable. Yet the vertically-implicit solution depends on the long time-step $\Delta t = 0.5$, as revealed by comparison with \code{ExplicitTimeDiscretization} using $\Delta t = 10^{-4}$.}
\label{fig:time-discretization}
\end{figure}

\code{VerticallyImplicitTimeDiscretization} permits longer time-steps when using fine vertical spacing.
Listing~\ref{list:time-discretization} illustrates the differences between vertically-implicit and explicit time discretization using one-dimensional tracer diffusion with a top-hat diffusivity profile.
The results are shown in figure~\ref{fig:time-discretization}.

\begin{jllisting}[float, caption={Diffusion of an initially linear tracer profile with a top-hat diffusivity. The three simulations shown in figure~\ref{fig:time-discretization} differ only in their time step and time discretization.}, label={list:time-discretization}]
using Oceananigans
using Oceananigans.TurbulenceClosures:
    ExplicitTimeDiscretization, VerticallyImplicitTimeDiscretization

function tracer_diffusion(Δt, time_discretization)
    grid = RectilinearGrid(size=20, z=(-2, 2), topology=(Flat, Flat, Bounded))
    κ(z, t) = abs(z) < 1
    closure = VerticalScalarDiffusivity(time_discretization; κ)
    model = HydrostaticFreeSurfaceModel(grid; closure, tracers=:c,
    velocities=PrescribedVelocityFields())
    set!(model, c=z -> z)
    simulation = Simulation(model; Δt, stop_time=0.5)
    run!(simulation)
    return model.tracers.c
end

explicit_large = tracer_diffusion(0.5, ExplicitTimeDiscretization())
explicit_small = tracer_diffusion(1e-4, ExplicitTimeDiscretization())
implicit_large = tracer_diffusion(0.5, VerticallyImplicitTimeDiscretization())
\end{jllisting}

\subsection{The pressure correction method for momentum in \code{NonhydrostaticModel}}
\label{fractional-step-method}

The \code{NonhydrostaticModel} uses a pressure correction method for the momentum equation~\eqref{nonhydrotatic-momentum-equations} that ensures $\bnabla \bcdot \bu = 0$.
We rewrite~\eqref{nonhydrotatic-momentum-equations} as 
\beq \label{momentum-tendency-definition}
\d_t \bu = - \bnabla p + b \bzh + \b{R}_u + \d_z \left ( \nu_z \d_z \bu \right ) \com
\eeq
where, if specified, $\nu_z$ is the vertical component of the viscosity that will be treated with a vertically-implicit time discretization, $\bnabla p$ is the total pressure gradient, and $\b{R}_u$ contains the remaining momentum tendency terms, excluding pressure, buoyancy, and vertically implicit diffusion.
We decompose $p$ into a ``hydrostatic anomaly'' $p_{hy}$ tied to the density anomaly $\rho'$, and a nonhydrostatic component $p_{nh}$, such that
\beq \label{pressure-decomposition}
p = p_{nh} + p_{hy} \com
\qquad \text{where} \qquad
\d_z p_{hy} \defn b \per
\eeq
Hydrostatic balance cancels the explicit buoyancy term, so that $-\bnabla p + b\bzh = -\bnabla p_{nh} - \nablah p_{hy}$.
We define $\b{G}_u \defn \b{R}_u - \nablah p_{hy}$, retaining the horizontal hydrostatic pressure gradient in the tendency used for the predictor.
Next, integrating~\eqref{momentum-tendency-definition} in time from $t^m$ to $t^{m+1}$ yields
\beq \label{fully-explicit-time-discretization}
\bu^{m+1} = \bu^m + \int_{t^m}^{t^{m+1}} \left [ \b{G}_u - \bnabla p_{nh} + \d_z \left ( \nu_z \d_z \bu \right ) \right ] \di t \per
\eeq
Next we introduce the predictor velocity $\tilde{\bu}$, defined such that
\beq \label{predictor-velocity}
\left ( 1 - \Delta t \, \d_z \nu_z^m \d_z \right ) \tilde{\bu} = \bu^m + \int_{t^m}^{t^{m+1}} \b{G}_u \di t \com
\eeq
or in other words, defined as a velocity-like field that cannot feel nonhydrostatic pressure gradient $\bnabla p_{nh}$.
Equation~\eqref{predictor-velocity} uses a semi-implicit treatment of vertical momentum diffusion which is similar but slightly different to the treatment of tracer diffusion in~\eqref{tracer-time-discretization},
\beq \label{semi-implicit-predictor-diffusion}
\int_{t^m}^{t^{m+1}} \d_z \left ( \nu_z \d_z \bu \right ) \di t \approx \Delta t \, \d_z \left ( \nu_z^m \d_z \tilde{\bu} \right ) \per
\eeq
The integral in \eqref{predictor-velocity} is evaluated using the RK3 scheme in \eqref{rk3}, as for tracers.
With a second-order discretization of vertical momentum diffusion, the predictor velocity in \eqref{predictor-velocity} may be computed with a tridiagonal solver.

Introducing a fully-implicit time discretization for $p_{nh}$,
\beq \label{implicit-pressure}
\int_{t^m}^{t^{m+1}} \bnabla p_{nh} \di t \approx \Delta t \, \bnabla p_{nh}^{m+1} \com
\eeq
and inserting \eqref{implicit-pressure} into \eqref{predictor-velocity}, we derive the pressure correction to the predictor velocity,
\beq \label{pressure-correction}
\bu^{m+1} - \tilde{\bu} = - \Delta t \, \bnabla p_{nh}^{m+1} \per
\eeq

The final ingredient needed to complete the pressure correction scheme is an equation for the nonhydrostatic pressure $p_{nh}^{m+1}$.
For this we form $\bnabla \bcdot \eqref{pressure-correction}$ and use $\bnabla \bcdot \bu^{m+1} = 0$ to obtain a Poisson equation for $p_{nh}^{m+1}$,
\beq \label{pressure-poisson}
\nabla^2 p_{nh}^{m+1} = \frac{\bnabla \bcdot \tilde{\bu}}{\Delta t} \per
\eeq
Boundary conditions for equation~\eqref{pressure-poisson} may be derived by evaluating $\bnh \bcdot \eqref{fully-explicit-time-discretization}$ on the boundary of the domain.

On a \code{RectilinearGrid}, we solve \eqref{pressure-poisson}
using an eigenfunction expansion of the discrete second-order Poisson operator $\nabla^2$ evaluated via the fast Fourier transform (FFT) in equispaced directions \cite{schumann1988fast} plus a tridiagonal solve in variably-spaced directions.
With the FFT-based solver, boundary conditions on $p_{nh}^{m+1}$ are accounted for by enforcing $\bnh \bcdot \tilde{\bu} = \bnh \bcdot \bu^{m+1}$ on boundary cells --- which is additional and separate from the definition of $\tilde{\bu}$ in \eqref{predictor-velocity}.
This alteration of $\tilde{\bu}$ on the boundary implicitly contributes the appropriate terms that account for inhomogeneous boundary-normal pressure gradients $\bnh \bcdot \bnabla p_{nh}^{m+1} \ne 0$ to the right-hand-side of \eqref{pressure-poisson} during the computation of $\bnabla \bcdot \tilde{\bu}$.

A preconditioned conjugate gradient iteration may be used on non-rectilinear grids, including complex domains.
For domains that immerse an irregular boundary within a \code{RectilinearGrid}, we have implemented an efficient, rapidly-converging preconditioner that leverages the FFT-based solver with masking applied to immersed cells.
The FFT-based preconditioner for solving the Poisson equation in irregular domains will be described in a forthcoming paper.

\subsection{Time discretization of the \code{HydrostaticFreeSurfaceModel}}

The \code{HydrostaticFreeSurfaceModel} uses a geopotential ($z$ or $z^\star$) vertical coordinate and may be integrated in time using either a fully \code{ExplicitFreeSurface}, an \code{ImplicitFreeSurface} utilizing a two-dimensional elliptical solve, or a \code{SplitExplicitFreeSurface}. The \code{SplitExplicitFreeSurface} supports both a linear free surface formulation and a nonlinear free surface formulation with the $z^\star$ free-surface-following vertical coordinate \cite{adcroft2004rescaled}; the latter is used for the global simulation shown in section~\ref{sec:clima-ocean} and described in detail by \citeA{silvestri2025gpu}.
For brevity, we describe here only the \code{SplitExplicitFreeSurface}, which is the most generally useful method.
The \code{SplitExplicitFreeSurface} substeps the depth-integrated or ``barotropic'' horizontal velocity $\bU_h$ along with the free surface displacement $\eta$ using a short time step while the depth-dependent, ``baroclinic'' velocities, along with tracers, are relatively stationary.

The barotropic horizontal transport $\bU_h$ is defined
\beq
\bU_h \defn \int_{-H}^\eta \bu_h \di z \com
\eeq
where $\bu_h = (u, v)$ is the total horizontal velocity and $H$ is the depth of the fluid.

Similarly integrating the horizontal momentum equations~\eqref{hydrostatic-equations} from $z=-H$ to $z=\eta$ yields an evolution equation for $\bU_h$,
\beq \label{momentum-equation-vertically-integrated}
\d_t \bU_h = - g (H + \eta) \nablah \eta + \int_{-H}^\eta \b{G}_{uh} \di z \com
\eeq
where $\b{G}_{uh}$ includes all the tendency terms that evolve ``slowly'' compared to the barotropic mode:
\beq \label{}
\b{G}_{uh} = -\nablah p - (\bu \bcdot \bnabla) \bu_h - \b{f} \times \bu - \bnabla \bcdot \b{\tau} + \b{F}_h \per
\eeq
Integrating continuity over the water column, using an impermeable bottom and the surface kinematic condition $\d_t\eta + \bu_h|_{z=\eta}\bcdot\nablah\eta = w|_{z=\eta}$, gives
\beq \label{free-surface-equation}
\d_t \eta = - \nablah \bcdot \bU_h \per
\eeq
The pair of equations \eqref{momentum-equation-vertically-integrated} and \eqref{free-surface-equation} characterize the evolution of the barotropic mode, which involves faster time-scales than the baroclinic mode evolution described by equations~\eqref{hydrostatic-equations}. 
To resolve both modes, we use a split-explicit algorithm where the barotropic mode is advanced in time using a smaller time-step than the one used for three-dimensional baroclinic variables. 
The three-dimensional variables are advanced with a Runge-Kutta scheme coupled to the barotropic substeps.
The coupling keeps the depth-integrated three-dimensional velocity consistent with the barotropic transport.
The Runge-Kutta stages and barotropic--baroclinic coupling are described by \citeA{silvestri2026rungekutta}.

\section{Finite volume spatial discretization}
\label{sec:finite-volume}

\newcommand{\V}{\mathscr{V}}
\newcommand{\A}{\mathscr{A}}

\begin{figure}[htp]
\centering
\includegraphics[width=\textwidth]{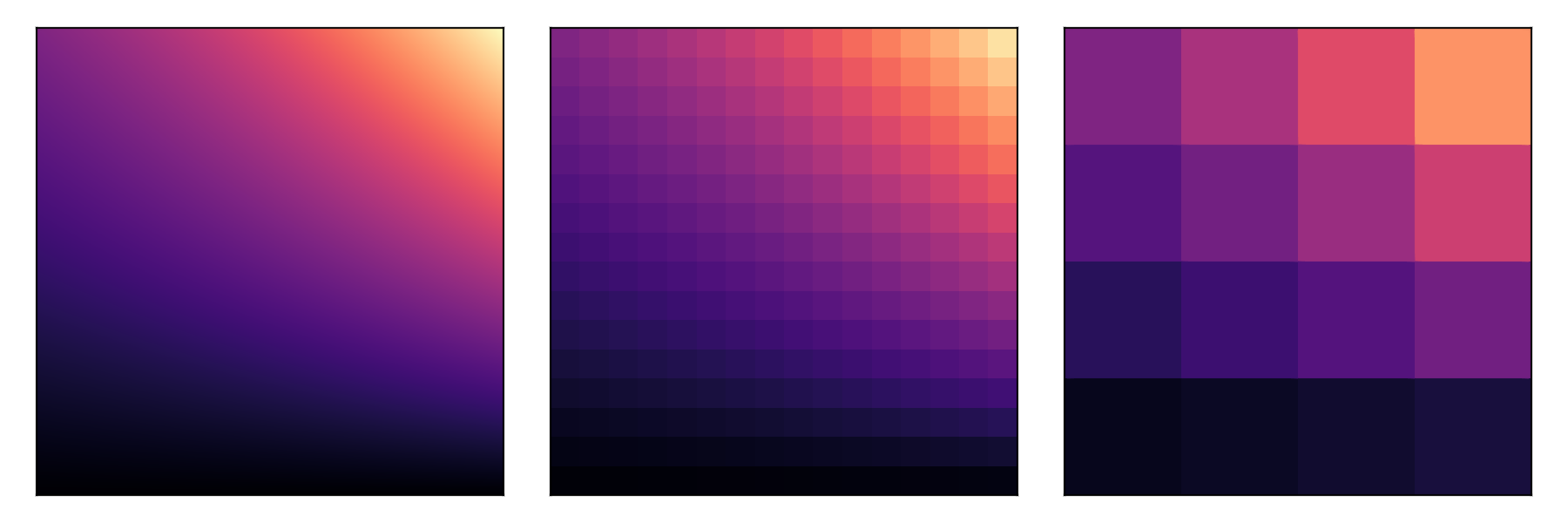}
\caption{Finite volume discretization of $\ee^{x} y$ on the unit square using $1024 \times 1024$, $16 \times 16$, and $4 \times 4$ cells.}
\label{fig:finite-volumes}
\end{figure}

Oceananigans uses a finite volume method in which fields are represented discretely by their average value over small local regions or ``finite volumes'' of the domain.
Listing~\ref{list:finite-volumes} discretizes $c = \ee^{x} y$ on three different grids that cover the unit square.
\begin{jllisting}[float, caption={Finite volume discretization of $\ee^x y$ on three grids over the unit square. The fields are visualized in figure~\ref{fig:finite-volumes}. The meaning of the ``Center'' in ``\code{CenterField}'' is discussed below.}, label={list:finite-volumes}]
topology = (Bounded, Bounded, Flat)
x = y = (0, 1)
c(x, y) = exp(x) * y

fine_grid = RectilinearGrid(size=(1024, 1024); x, y, topology)
c_fine = CenterField(fine_grid)
set!(c_fine, c)

medium_grid = RectilinearGrid(size=(16, 16); x, y, topology)
c_medium = CenterField(medium_grid)
regrid!(c_medium, c_fine)

coarse_grid = RectilinearGrid(size=(4, 4); x, y, topology)
c_coarse = CenterField(coarse_grid)
regrid!(c_coarse, c_medium)
\end{jllisting}
On the finest grid, each cell-averaged value $c_{ij}^\mathrm{fine}$ is computed approximately using \code{set!} to evaluate $\ee^{x} y$ at the center of each finite volume, where $i, j$ denote the $x$ and $y$ indices of the finite volumes.
On the medium and coarse grids, the cell averages are computed by averaging or ``regridding'' fields from the next finer grid.
This computation produces three fields with identical integrals over the unit square.
For example, integrals are computed exactly by summing discrete fields over all cells,
\beq
\int c \di x \di y = \sum_{i, j}^{1024, 1024} \V^\mathrm{fine}_{ij} c^\mathrm{fine}_{ij} = \sum_{i, j}^{16, 16} \V^\mathrm{medium}_{ij} c^\mathrm{medium}_{ij} = \sum_{i, j}^{4, 4} \V^\mathrm{coarse}_{ij} c^\mathrm{coarse}_{ij} \com
\eeq
where $\V_{ij}$ is the ``volume'' of the cell with indices $i, j$ (more accurately an ``area'' in this two-dimensional situation).
Figure~\ref{fig:finite-volumes} visualizes the three fields.

\begin{figure}[htp]
\centering
\includegraphics[width=\textwidth]{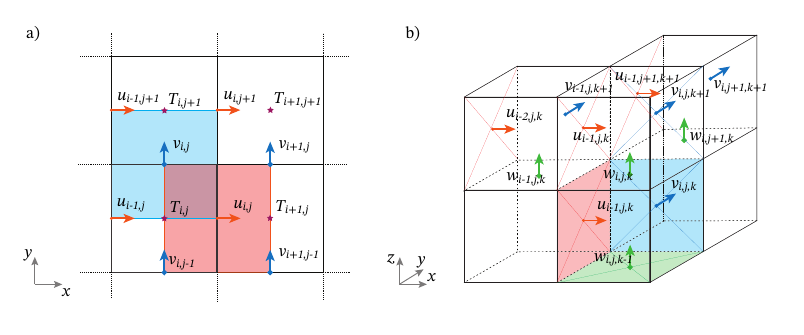} %
\caption{Locations of cell centers and interfaces on a two-dimensional (a) and three-dimensional (b) staggered grid. In (a), the red and blue shaded regions highlight the volumes in the dual $u$-grid and $v$-grid, located at \code{(Face, Center, Center)} and \code{(Center, Face, Center)}, respectively. In (b), the shaded regions highlight the facial areas used in the fluxes computations, denoted with $\A_x$, $\A_y$, and $\A_z$.}
\label{fig:staggered-grid}
\end{figure}

The discrete calculus and arithmetic operations required to solve the governing equations of the \code{NonhydrostaticModel} and \code{HydrostaticFreeSurfaceModel} use the system of ``staggered grids'' described by \citeA{arakawa1977computational}.
Both models use ``C-grid'' staggering, where cells for tracers, pressure, and the divergence of the velocity field $\bnabla \bcdot \bu$ are co-located, and cells for velocity components $\bu = (u, v, w)$ are staggered by half a cell width in the $x$-, $y$-, and $z$-directions, respectively.
Listing~\ref{list:staggered-grid} illustrates grid construction and notation for a one-dimensional staggered grid with unevenly-spaced cells.
Figure~\ref{fig:staggered-grid} visualizes 2- and 3-dimensional staggered grids, indicating the location of certain variables.

\begin{jllisting}[caption={A one-dimensional staggered grid.}, label={list:staggered-grid}]
using Oceananigans

grid=RectilinearGrid(topology=(Bounded, Flat, Flat), size=4, x=[0, 0.2, 0.3, 0.7, 1])

u = Field{Face, Center, Center}(grid)
c = Field{Center, Center, Center}(grid)

xnodes(u)       # [0.0, 0.2, 0.3, 0.7, 1.0]
xnodes(c)       # [0.1, 0.25, 0.5, 0.85]
location(∂x(c)) # (Face, Center, Center)
\end{jllisting}

\subsection{A system of composable operators}

A convention for indexing is associated with staggered locations.
Face indices are ``left'' of cell indices.
This means that difference operators acting on fields at cells differ from those that act on face fields.
To illustrate this we introduce Oceananigans-like difference operators,
\begin{jllisting}
δxᶠᶜᶜ(i, j, k, grid, c) = c[i, j, k] - c[i-1, j, k]
δxᶜᶜᶜ(i, j, k, grid, u) = u[i+1, j, k] - u[i, j, k]
\end{jllisting}
where superscripts denote the location of the \textit{result} of the operation.
For example, the difference $\delta_x^\mathrm{fcc}$ acts on fields located at ccc (meaning cell \code{Center} in the $x$-, $y$-, and $z$-directions respectively).
Complementary to the difference operators are reconstruction of ``interpolation'' operators,
\begin{jllisting}
ℑxᶠᶜᶜ(i, j, k, grid, c) = (c[i, j, k] + c[i-1, j, k]) / 2
ℑxᶜᶜᶜ(i, j, k, grid, u) = (u[i+1, j, k] + u[i, j, k]) / 2
\end{jllisting}

The prefix arguments \code{(i, j, k, grid)} are more than convention: the prefix enables system for \textit{composing} operators.
For example, defining
\begin{jllisting}
δxᶠᶜᶜ(i, j, k, grid, f::Function, args...) =
    f(i, j, k, grid, args...) - f(i-1, j, k, grid, args...)
    
δxᶜᶜᶜ(i, j, k, grid, f::Function, args...) =
    f(i+1, j, k, grid, args...) - f(i, j, k, grid, args...)
\end{jllisting}
leads to a concise definition of the second-difference operator:
\begin{jllisting}
δ²xᶜᶜᶜ(i, j, k, grid, f::Function, a...) = δxᶜᶜᶜ(i, j, k, grid, δxᶠᶜᶜ, f, a...)
\end{jllisting}
Operator composition is used throughout Oceananigans source code to implement stencil operations.

\subsection{Discrete identities and kinetic energy conservation}
\label{app:discrete-identities}

Oceananigans' vector invariant schemes share a circulation-based vertical vorticity operator, but differ in the reconstruction and coupling of vorticity to velocity.
Using the difference operators and staggered locations defined above, the vertical curl is discretized as
\beq
    (\bnabla\times\bu)_z \approx
    \frac{1}{\A_z^{\mathrm{ffc}}}
    \left[\delta_x(\Delta y\,v)-\delta_y(\Delta x\,u)\right] \com
\eeq
where $\A_z^{\mathrm{ffc}}$ is the area enclosed by the circulation stencil, and the edge lengths $\Delta x$ and $\Delta y$ are evaluated at the corresponding velocity locations.
For the compatible gradient of a cell-centered scalar $\psi$, the edge-integrated components are $\Delta x\,u=\delta_x\psi$ and $\Delta y\,v=\delta_y\psi$.
The numerator of the discrete vertical curl of this gradient therefore vanishes:
\beq
    \delta_x\delta_y\psi-\delta_y\delta_x\psi=0 \per
\eeq
The other curl components follow the same pattern.

Kinetic energy conservation imposes a separate condition on the momentum advection scheme.
For example, the optional \code{EnergyConservingUpwinding} reconstruction pairs upwinded vorticity symmetrically between the horizontal momentum equations.
If $G_i^u$ and $G_j^v$ denote the vorticity contributions to the velocity tendencies, and $\V_i^{\mathrm{fcc}}$ and $\V_j^{\mathrm{cfc}}$ the corresponding control-volume weights, this pairing gives
\begin{equation}
    \left.\frac{\mathrm{d}K_h}{\mathrm{d}t}\right|_{\mathrm{vorticity}}
    = \sum_i \V_i^{\mathrm{fcc}} u_i G_i^u + \sum_j \V_j^{\mathrm{cfc}} v_j G_j^v = 0 ,
\end{equation}
where $K_h = \tfrac{1}{2}\sum_i \V_i^{\mathrm{fcc}} u_i^2 + \tfrac{1}{2}\sum_j \V_j^{\mathrm{cfc}} v_j^2$ is the integrated discrete horizontal kinetic energy.
The identity applies to a closed or periodic domain with boundary treatment that preserves the pairing.

\subsection{Tracer flux divergences, advection schemes, and reconstruction}

The divergence of a tracer flux $\b{J} = J_x \bxh + J_y \byh + J_z \bzh$ is discretized conservatively by the finite volume method via
\beq \label{flux-divergence}
\bnabla \bcdot \b{J} \approx
    \frac{1}{\V_c} \Big [
        \delta_x \big (\underbrace{\strut \A_x J_x}_{\mathrm{fcc}} \big ) + 
        \delta_y \big (\underbrace{\strut \A_y J_y}_{\mathrm{cfc}} \big ) + 
        \delta_z \big (\underbrace{\strut \A_z J_z}_{\mathrm{ccf}} \big )
    \Big ] \com
\eeq
where $\delta_x, \delta_y, \delta_z$ are difference operators in $x, y, z$, $\V_c$ denotes the volume of the tracer cells, $\A_x$, $\A_y$, and $\A_z$ denote the areas of the tracer cell faces with surface normals $\bxh$, $\byh$, and $\bzh$, respectively.
Equation~\eqref{flux-divergence} indicates the location of each flux component: fluxes into tracers cell at ccc are computed at the cell faces located at fcc, cfc, and ccf.

The advective tracer flux in~\eqref{nonhydrostatic-tracer-equation} is written in ``conservative form'' using incompressibility~\eqref{continuity}, and then discretized similarly to \eqref{flux-divergence} to form
\beq \label{advection-flux-divergence}
\bu \bcdot \bnabla c = \bnabla \bcdot \left ( \bu c \right ) \approx \frac{1}{\V_c} \Big [
    \delta_x \big (\A_x u \bigl \lfloor c \bigr \rceil_x \big ) + 
    \delta_y \big (\A_y v \bigl \lfloor c \bigr \rceil_y \big ) + 
    \delta_z \big (\A_z w \bigl \lfloor c \bigr \rceil_z \big ) \Big ] \com
\eeq
where $\bigl \lfloor c \bigr \rceil_x$ denotes a \textit{reconstruction} of $c$ in the $x$-direction from its native location ccc to the tracer cell interface at fcc;
$\bigl \lfloor c \bigr \rceil_y$ and $\bigl \lfloor c \bigr \rceil_z$ in \eqref{advection-flux-divergence} are defined similarly.

The advective fluxes $\bu c$ must be computed on interfaces between tracer cells, where the approximate value of $c$ must be reconstructed.
(Velocity components like $u$ must also be reconstructed on interfaces. Within the C-grid framework, we approximate $u$ on tracer cell interfaces directly using the values ${u}_{ijk}$, which represent $u$ averaged over a region encompassing the interface.)
The simplest kind of reconstruction is \code{Centered(order=2)}, which is equivalent to taking the average between adjacent cells,
\beq \label{centered-reconstruction}
\langle c \rangle_i = \half \left ( c_i + c_{i-1} \right ) \com
\eeq
where $\langle c \rangle_i$ denotes the centered reconstruction of $c$ on the interface at $x=x_{i-1/2}$.
Also in \eqref{centered-reconstruction} the $j, k$ indices are implied and we have suppressed the direction $x$ to lighten the notation.
Reconstruction stencils for \code{Centered(order=$N$)} are automatically generated for even $N$ up to $N_\mathrm{max}=12$, where $N_\mathrm{max}$ is an adjustable parameter in the source code.
All subsequent reconstructions are described in the $x$-direction only.

\code{Centered} schemes should be used when explicit dissipation justified by a \textit{physical} rationale dominates at the grid scale.
In scenarios where dissipation is needed solely for artificial reasons, we find applications for \code{UpwindBiased} schemes, which use an odd-order stencil biased against the direction of flow.
For example, \code{UpwindBiased(order=1)} and \code{UpwindBiased(order=3)} schemes are written
\beq \label{upwind-biased}
u [ c ]^1_x = \left \{
\begin{matrix}
    u \, c_{i-1} & \text{if } u > 0 \com \\
    u \, c_i     & \text{if } u < 0 \com    
\end{matrix} \right .
\quad \text{and} \quad
u [ c ]^3_x = \left \{
\begin{matrix}
    u \, \tfrac{1}{6} \left ( - c_{i-2} + 5 c_{i-1} + 2 c_i \right ) & \text{if } u > 0 \com \\
    u \, \tfrac{1}{6} \left ( 2 c_{i-1} + 5 c_i - c_{i+1} \right ) & \text{if } u < 0 \com   
\end{matrix} \right .
\eeq
where $[c]^N_x$ denotes $N^\mathrm{th}$-order upwind reconstruction in the $x$-direction.
(Note that $u [c]^N_x = 0$ if $u=0$.)

The compact form of equations \eqref{upwind-biased} demonstrates how upwind schemes introduce variance dissipation through numerical discretization. In particular, an \code{UpwindBiased(order=1)} reconstruction can be rewritten as a sum of a \code{Centered(order=2)} discrete advective flux and a discrete diffusive flux
\beq
u [ c ]^1_x = u \frac{c_{i} + c_{i-1}}{2} - \kappa_1 \frac{c_{i} - c_{i-1}}{\Delta x} \com \ \ \text{where} \ \ \ \kappa_1 = \frac{|u|\Delta x}{2} \per
\eeq
Reordering the \code{UpwindBiased(order=3)} advective flux in the same manner recovers a sum of a \code{Centered(order=4)} advective flux and a 4th-order hyperdiffusive flux, equivalent to a finite volume approximation of
\beq
uc + \kappa_3\frac{\partial^3 c}{\partial x^3} \com \ \ \text{where} \ \ \kappa_3 = \frac{|u|\Delta x^3}{12} \per
\eeq
\code{UpwindBiased} reconstruction can always be reordered to expose a sum of \code{Centered} reconstruction and a high-order diffusive flux with a velocity-dependent diffusivity.
The diffusive operator associated with \code{UpwindBiased(order=1)} and \code{UpwindBiased(order=3)} is enough to offset the dispersive errors of the \code{Centered} component and, therefore, eliminate the artificial explicit diffusion needed for stability purposes.
However, this approach does not scale to high order since the diffusive operator associated with a high order \code{UpwindBiased} scheme (5th, 7th, and so on), becomes quickly insufficient to eliminate spurious errors associated with the \code{Centered} component \cite{godunov1959difference}.

The inability to achieve high order and, therefore, low dissipation motivated 
the implementation of Weighted, Essentially Non-Oscillatory (WENO) reconstruction \cite{shu1997essentially, shu2009high}.
WENO is a non-linear reconstruction scheme that combines a set of odd-order linear reconstructions obtained by stencils that are shifted by a value $s$ relative to the canonical \code{UpwindBiased} stencil, using a weighting scheme for each stencil that depends on the smoothness of the reconstructed field $c$.
Since the constituent stencils are lower-order than the WENO order, this strategy yields a scheme whose order of accuracy adapts depending on the smoothness of the reconstructed field.
In smooth regions high-order is retained, while the order quickly decreases in the presence of noisy regions, decreasing the order of the associated diffusive operator.
WENO proves especially useful for turbulence-resolving simulations (either at meter or planetary scales) without requiring any additional explicit artificial dissipation \cite{pressel2017numerics, silvestri2024new}.

To illustrate how WENO works we consider a fifth-order WENO scheme for $u>0$,
\beq \label{fifth-order-weno}
\{ c \}^5 = \gamma_{-1} [c]^{3, -1} + \gamma_0 [c]^{3, 0} + \gamma_1 [c]^{3, 1} \com
\eeq
where the notation $[c]^{3, s}$ denotes an \code{UpwindBiased} stencil \textit{shifted} by $s$ indices, such that $[c]^3 \defn [c]^{3, 0}$.
The shifted upwind stencils $[c]^{N, s}_i$ evaluated at index $i$ are defined
\beq
[c]^{3, s}_i = \frac{1}{6} \left \{
\begin{matrix}
2 c_{i-1} + 5 c_{i} - c_{i+1}      & \quad \text{for } s = -1 \com \\
- c_{i-2} + 5 c_{i-1} + 2 c_{i}    & \quad \text{for } s = 0 \com \\
2 c_{i-3} - 7 c_{i-2} + 11 c_{i-1} & \quad \text{for } s = 1 \per
\end{matrix} \right .
\eeq
The nonnegative weights $\gamma_s(c)$, with $s=-1,0,1$ and $\sum_s \gamma_s=1$, are determined by a smoothness metric that produces $\{c\}^5 \approx [c]^5$ when $c$ is smooth, but limits to the more diffusive $\{c\}^5 \approx [c]^3$ when $c$ changes abruptly.
Thus WENO adaptively introduces dissipation as needed based on the smoothness of $c$, yielding stable simulations with a high effective resolution that require no artificial dissipation.
WENO can alternatively be interpreted as adding an implicit hyperviscosity that adapts from low- to high-order depending on the local nature of the solution.
To compute the weights $\gamma_s(c)$, we use the WENO-Z formulation \cite{balsara2000monotonicity}.

\begin{figure}[htp]
\centering
\includegraphics[width=1.0\textwidth]{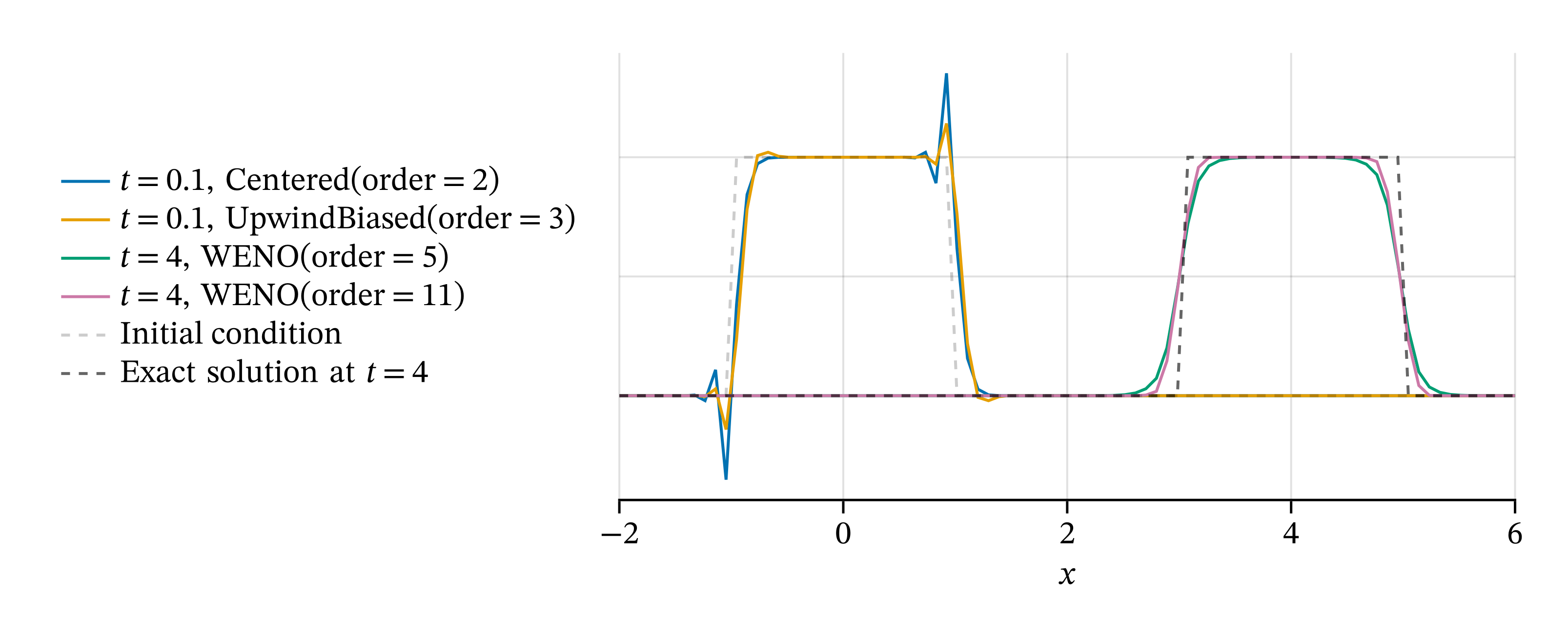}
\caption{Advection of a top hat tracer distribution in one dimension using various advection schemes. \code{Centered} and \code{UpwindBiased} schemes produce oscillations at sharp gradients, while WENO adapts its reconstruction to minimize both oscillations and dissipation.}
\label{fig:one-dimensional-advection}
\end{figure}

The properties of \code{Centered}, \code{UpwindBiased}, and WENO reconstruction are investigated by listing~\ref{list:one-dimensional-advection}, which simulates the advection of a top hat tracer distribution.
The results are plotted in figure~\ref{fig:one-dimensional-advection}.

\begin{jllisting}[caption={A script that advects a top hat tracer profile in one-dimension with a constant prescribed velocity. We use \code{halo=6} to accommodate schemes up to \code{WENO(order=11)}.}, label={list:one-dimensional-advection}]
using Oceananigans

grid=RectilinearGrid(size=128; x=(-4, 8), halo=6, topology=(Periodic, Flat, Flat))
advection = WENO(order=9) # Centered(order=2), UpwindBiased(order=3)
velocities = PrescribedVelocityFields(u=1)
model = HydrostaticFreeSurfaceModel(; grid, velocities, advection, tracers=:c)

top_hat(x) = abs(x) > 1 ? 0 : 1
set!(model, c = top_hat)

simulation = Simulation(model, Δt=1/grid.Nx, stop_time=4)
run!(simulation)    
\end{jllisting}

\subsubsection{Discretization of momentum advection}

The discretization of momentum advection with a flux form similar to \eqref{advection-flux-divergence} is more complex than the tracer case because both the advecting velocity and advected velocity require reconstruction.
We use the method described by \citeA{ghosh2012high} and \citeA{pressel2015large}, wherein advecting velocities are constructed with a high-order \code{Centered} scheme when the advected velocity component is reconstructed with a high-order \code{UpwindBiased} or WENO scheme.
We have also developed a novel WENO-based method for discretizing momentum advection in the rotational or ``vector invariant'' form especially appropriate for representing mesoscale and submesoscale turbulent advection on curvilinear grids \cite{silvestri2024new}.

\section{Parallelization}
\label{app:parallelization}

\begin{figure}
    \centering
    \includegraphics[width=\textwidth]{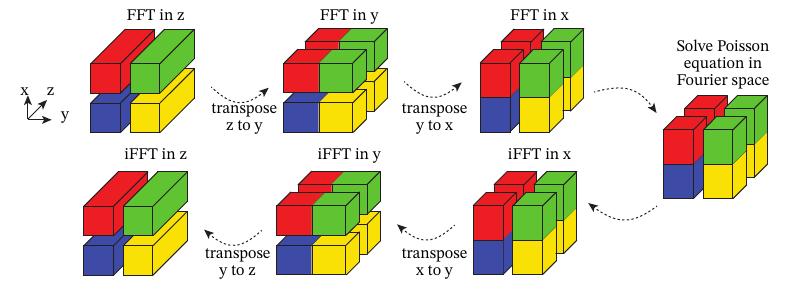}
    \caption{A schematic showing the distributed Poisson solver procedure with a pencil parallelization that divides the domain in two ranks in both~$x$ and~$y$. The schematic highlights the data layout in the ranks during each operation.}
    \label{fig:transpose}
\end{figure}

Oceananigans supports distributed computations with slab and pencil domain decomposition. The interior domain is extended using ``halo'' or ``ghost'' cells that hold the results of interprocessor boundaries. ``halo'' cells are updated before the computation of tendencies through asynchronous send / receive operations using the message passing interface (MPI) Julia library \cite{byrne2021mpi}. For a detailed description of the parallelization strategy of the \code{HydrostaticFreeSurfaceModel}; see \citeA{silvestri2025gpu}. The \code{NonhydrostaticModel} implements the same overlap of communication and computation for halo exchange before the calculation of tendencies.
For the FFT-based three-dimensional pressure solver, we implement a transpose algorithm that switches between $x$-local, $y$-local, and $z$-local configurations to compute efficiently the discrete transforms.
The transpose algorithm for the distributed FFT solver is shown in figure~\ref{fig:transpose}.

\FloatBarrier
\section{Table of numerical examples}

Table~\ref{table:numerical-experiments} provides a list of the numerical experiments in this paper.

\begin{table}
\caption{DNS: Direct Numerical Simulation. LES: Large Eddy Simulation.}
\begin{center}
\label{table:numerical-experiments}
\begin{tabular}{ l | l | l }
 Description & Code & Visualization \\ 
 \hline
2D turbulence using \code{WENO(order=9)} advection               & listing~\ref{list:first-script}              & fig.~\ref{fig:first-impression} \\
2D turbulence with moving tracer source                   & listing~\ref{list:moving-source}             & fig.~\ref{fig:first-impression} \\
DNS and LES of flow around a cylinder at various $Re$     & listing~\ref{list:flow-around-cylinder}      & fig.~\ref{fig:flow-around-cylinder} \\
DNS of cabbeling in freshwater                            & listing~\ref{list:cabbeling}                 & fig.~\ref{fig:cabbeling} \\
LES of the Eady problem with \code{WENO(order=9)}                & listing~\ref{list:eady-les}                  & fig.~\ref{fig:eady-les} \\
Tidally-oscillating flow past Three Tree Point            & listing~\ref{list:headland-les}              & fig.~\ref{fig:headland} \\
Internal waves generated by tidal forcing over bathymetry & listing~\ref{list:internal-tide}             & fig.~\ref{fig:internal-tide} \\
Comparison of vertical mixing parameterizations           & listing~\ref{list:parameterizations}         & fig.~\ref{fig:parameterizations} \\
Global and near-global baroclinic instability on four grids         & listing~\ref{list:near-global}               & fig.~\ref{fig:near-global} \\
Realistic ocean simulation with NumericalEarth             & listing~\ref{list:clima-ocean}               & fig.~\ref{fig:clima-ocean} \\
Tracer diffusion with various time discretizations        & listing~\ref{list:time-discretization}       & fig.~\ref{fig:time-discretization} \\
Visualization of the finite volume discretization         & listing~\ref{list:finite-volumes}            & fig.~\ref{fig:finite-volumes} \\
One-dimensional advection of a top-hat tracer profile     & listing~\ref{list:one-dimensional-advection} & fig.~\ref{fig:one-dimensional-advection}
\end{tabular}
\end{center}
\end{table}

\FloatBarrier
\section{Acronyms and abbreviations}
\label{sec:acronyms}

Table~\ref{table:acronyms} defines acronyms and abbreviations used in the text, captions, and code examples. Software and dataset names without a standard expansion are described by their purpose.

\begin{longtable}{@{}p{0.19\textwidth}p{0.75\textwidth}@{}}
\caption{Acronyms, abbreviations, and abbreviated software and dataset names.}
\label{table:acronyms}\\
\toprule
Abbreviation & Meaning \\
\midrule
\endfirsthead
\multicolumn{2}{l}{Table~\thetable\ (continued)}\\
\toprule
Abbreviation & Meaning \\
\midrule
\endhead
\midrule
\multicolumn{2}{r}{Continued on next page}\\
\endfoot
\bottomrule
\endlastfoot
2D, 3D & Two-dimensional, three-dimensional \\
ACE & Ai2 Climate Emulator \\
AI & Artificial intelligence \\
AMReX & Software framework for block-structured adaptive mesh refinement \\
CATKE & Convective Adjustment and Turbulent Kinetic Energy \\
CFL & Courant--Friedrichs--Lewy \\
CPU & Central processing unit \\
CROCO & Coastal and Regional Ocean Community model \\
CUDA & NVIDIA's GPU computing platform; accessed through CUDA.jl in the examples \\
DNS & Direct numerical simulation \\
DSL & Domain-specific language \\
ECCO & Estimating the Circulation and Climate of the Ocean \\
EN4 & Met Office dataset of quality-controlled ocean temperature and salinity profiles and objective analyses, version 4 \\
ETOPO1, ETOPO2022 & Global topography and bathymetry products: the 1-arc-minute ETOPO1 product and the ETOPO 2022 product \\
FFT & Fast Fourier transform \\
GLORYS & Global Ocean Reanalyses and Simulations \\
GM & Gent--McWilliams \\
GPU & Graphics processing unit \\
I/O & Input/output \\
JAX & Python library for array computation and automatic differentiation \\
JRA55 & Japanese 55-year Reanalysis (also written JRA-55) \\
JRA55-do & JRA-55-based surface dataset for driving ocean--sea-ice models \\
LES & Large eddy simulation \\
MITgcm & Massachusetts Institute of Technology general circulation model \\
ML & Machine learning \\
MOM6 & Modular Ocean Model, version 6 (used by Regional-mom6) \\
MPI & Message Passing Interface \\
PiCLES & Particle-in-Cell for Efficient Swell \\
REMORA & Regional Model of the Ocean Refined Adaptively \\
RK3 & Third-order Runge--Kutta time-stepping scheme \\
ROMS & Regional Ocean Modeling System \\
SDPD & Simulated days per wall-clock day \\
SYPD & Simulated years per wall-clock day \\
TEOS-10 & Thermodynamic Equation of Seawater 2010 \\
TKE & Turbulent kinetic energy \\
TPU & Tensor processing unit \\
WENO & Weighted essentially non-oscillatory \\
WENO-Z & WENO reconstruction with the Z weighting formulation \\
\end{longtable}

\FloatBarrier
\section*{Open Research Section}

Oceananigans is available at the GitHub repository \href{https://github.com/CliMA/Oceananigans.jl}{\nolinkurl{github.com/CliMA/Oceananigans.jl}} and
NumericalEarth is available at \href{https://github.com/NumericalEarth/NumericalEarth.jl}{\nolinkurl{github.com/NumericalEarth/NumericalEarth.jl}}.
Oceananigans documentation lives at \href{https://clima.github.io/OceananigansDocumentation}{\nolinkurl{clima.github.io/OceananigansDocumentation}}.
All the scripts that reproduce the simulations and figures in this paper are available at the GitHub repository \href{https://github.com/glwagner/OceananigansPaper}{\nolinkurl{github.com/glwagner/OceananigansPaper}}.
Visualizations were made using Makie.jl \cite{Makie.jl}.

\section*{Acknowledgments}
We thank all six reviewers, Stephen Griffies, and Florian Lemarié for their time and constructive feedback, which helped improve this manuscript.
This project is supported by Schmidt Sciences, LLC and by the National Science Foundation grant AGS-1835576.
N.C.C.~is additionally supported by the Australian Research Council under the Center of Excellence for the Weather of the 21st Century CE230100012 and the Discovery Project DP240101274. AI was used to assist with editing, code, and figure preparation.

\noindent {\bf Conflict of interest}

\noindent The authors have no conflicts of interest to disclose.

\bibliography{refs.bib}

\end{document}